\documentclass[twocolumn]{aa} 
\usepackage{natbib}
\usepackage{graphicx}
\usepackage{rotating}
\usepackage{txfonts}

\begin{document}

\title{Loops formed by tidal tails as fossil records of a major merger}

   \author{Jianling Wang
          \inst{1,2}
          \and
          Francois Hammer\inst{1}
          \and
	  E. Athanassoula\inst{3}
          \and
          Mathieu Puech\inst{1}
          \and
          Yanbin Yang\inst{1,2}
        \and
         Hector Flores\inst{1}
          }

   \institute{Laboratoire GEPI, Observatoire de Paris, CNRS-UMR8111, 
		Univ. Paris-Diderot, 5 place Jules Janssen, 92195 Meudon France                                                          
              \email{jianling.wang@obspm.fr}  \email{francois.hammer@obspm.fr}
	 \and
		NAOC, Chinese Academy of Sciences, A20 Datun Road, 100012 Beijing, China 
             \and
             Laboratoire d'Astrophysique de Marseille, Observatoire Astronomique de Marseille Provence, Technop\^ole de Etoile,
Site de Chateau-Gombert, 38 rue Fr\'ed\'eric Joliot-Curie, 13388 Marseille C\'edex 13, France
             }

   \date{Received ; accepted }

 
  \abstract
   {
 Many haloes of nearby disc galaxies contain
 faint and extended features, including loops, which are often interpreted as relics of satellite infall in
 the main galaxy's potential well. In most cases, however, the residual
 nucleus of the satellite is not seen, although it is predicted by
 numerical simulations.
   }
   {
We test whether such faint and extended features can be associated to
gas-rich, major mergers, which may also lead to disc
rebuilding and thus be a corner stone for the formation of spiral
galaxies. 
Our goal is to test whether the major merger scenario can
provide a good model for a particularly difficult case, that of NGC
5907, and to compare to the scenario of a satellite infall.
   }
   {
   Using the TreeSPH code GADGET-2, we model the formation of an
     almost bulge-less galaxy similar to NGC 5907 (B/T $\le$ 0.2) after a gas-rich major merger. 
    First, we trace tidal tail particles captured by the galaxy gravitational potential  to verify
    whether they can form loops similar to those discovered in the
    galactic haloes. 
   }
   { 
   We indeed find that 3:1 major mergers can
     form features similar to the loops found in many galactic haloes,
     including in NGC 5907, and can reproduce
     an extended thin disc, a bulge, as well as the pronounced warp of the gaseous disc. Relatively small bulge fractions can be reproduced by a
     large gas fraction in the progenitors, as well as appropriate
     orbital parameters.    
     }
   { 
   Even though it remains difficult to fully cover the large volume
   of free parameters, the present modelling of the loops in NGC 5907 proves  
   that they could well be the result of a major merger. It has many
   advantages over the satellite infall scenario; e.g., it solves the problem of 
   the visibility of the satellite remnant, and it may explain some
   additional features in the NGC 5907 halo, as well as some gas
   properties of this system. For orbital parameters derived from cosmological simulations, the loops in NGC 5907 can be reproduced by major mergers (3:1 to 5:1) and possibly by intermediate mergers (5:1 to 12:1). 
   The major merger scenario thus challenges 
   the minor merger one and could explain many properties that
   haloes of spiral galaxies have in common, including their red colours
   and the presence of faint extended features. 
   }


\maketitle

%

\section{Introduction}

The hierarchical galaxy formation theory assumes that galaxies have been assembled through mergers 
of smaller entities. The relative impact of both minor and major mergers is still being intensively debated 
\citep{fm2008,lin2004,lin2008,hopkins2008,hopkins2009,bell2006,lotz2008,bridge2007}
Peculiar galaxies are mostly related to merger or interaction events. For example, most
luminous infrared galaxies \citep[LIRGs;][]{sm1996} 
are merging or interacting galaxies, as revealed by their detailed
substructures with multiple nuclei \citep[see][]{haan2011}. \citet{toomre1977}
first pointed out that a major merger could result in the formation of
elliptical galaxies. Later simulations showed that major mergers of
gas-poor progenitors will destroy the disc and form a galaxy with a large
spheroidal component.  

Partly due to their frequency, even minor mergers challenge the
survival of discs \citep{to1992,stewart2009}. Thus
the number of unperturbed discs at low redshift argues that 
there should be some mechanisms that either preserve the disc from
being destroyed or allow it to be rebuilt after the 
merger event. Analyses of merger remnants show that discs can
survive from a merger. Simulations with low gas fractions 
($\le$ 10\%) also show that a significant fraction of the gas can survive a 
major merger and form a new disc \citep{bh1996,barnes2002}. 
 By observing and analysing the progenitors of local spirals at
 intermediate redshift, \citet{hammer2005,hammer2007,hammer2009} 
suggest that many present-day discs could have been rebuilt after a major merger. Recent theoretical 
advances also support this scenario. Many hydrodynamical 
simulations indicate that new discs can be rebuilt 
after a major merger between disc galaxies \citep{robertson2006,hopkins2009,sh2005},
provided the initial
gas fraction is sufficiently high. In this process gas 
plays the key role in the disc rebuilding process. High redshift
galaxies have much higher gas fraction than local spirals, 
consistently with the above scenario. 
 
It is expected that about half of the local disc galaxies have
experienced a major merger in the past 9 Gyr \citep{puech2011},
and it should thus be 
expected to find imprints of such events within the remnant halo. 
For example M31, the largest spiral of the Local Group, contains many
faint structures in its halo. The Giant 
Stream \citep{ibata2001} is the most famous one, together with the large north-east loop with metal-poor stars recently discovered by the PANDA team 
\citep{richardson2011}. Most of these 
faint features in the halo of M31 have often been thought to be caused by numerous minor merger events, as many as the stream 
number. \citet{hammer2010}, however, show that most of these faint
features could, instead, be related to a single major merger event. 
 In their models, most 
of the faint structures, including the Giant Stream, are due to stars coming back from tidal
tails, a process that can be maintained for several Gyr.  

If the disc of many spirals is indeed rebuilt after a major merger, it is expected that tidal tails can be a fossil record and that
there should be many loops and streams in their haloes. Recently \citet{m2010} have conducted a pilot survey of 
isolated spiral galaxies in the Local Volume up to a low surface
brightness sensitivity of $\sim 28.5$ mag/arcsec$^2$ in $V$ band. 
They find that many of these galaxies have loops or streams of
various shapes and interpret these structures as evidence 
 of minor merger or satellite infall. However, if these loops are caused by minor mergers, the residual of the satellite core should be detected
according to numerical simulations. Why is it hardly ever detected ?

The above question is the starting point for this paper, which intends
to test whether a major merger scenario could, or could not,  
reproduce the observations of faint loops in nearby galaxy haloes
better than the minor merger one. We choose to study here the NGC5907 galaxy and its faint loop 
system observed by \citet[hereafter M08]{m2008}
because it is often regarded as the best evidence of a minor satellite interaction with a late-type spiral,
 i.e. not an easy configuration to be reproduced by a major merger.

The paper is organised as follows. In Sect. 
\ref{sec:property} we describe the properties of NGC 5907 and Sect. \ref{sec:haloproperty} summarises the peculiar features found in its halo.  Simulation methods and initial conditions are 
described in Section \ref{sec:simu}, and in Section 
\ref{sec:result} we present the results of our simulations of both the galaxy and the associated loops.  In Section 
\ref{sec:discuss} we discuss whether a major merger can reproduce a galaxy with similar properties to NGC 5907, and we then compare the relative merits of major and minor mergers. In appendices, we present the implementation of star formation in the GADGET-2 code (Appendix A) and we give a general description of the loop properties formed during  a major merger (Appendix B).

\section{Properties of the NGC 5907 galaxy} 
\label{sec:property}

NGC 5907 is a nearby Sc type spiral. There is a large uncertainty in the distance of NGC 5907 due to its peculiar motion. 
Zepf et al. (2000) got a distance 13.5 $\pm$ 2.1 Mpc from a combination of the Tully-Fisher relation in both optical and near infrared 
and a peculiar motion model, and they adopted a round value 14 Mpc, which is used by M08. For consistency, we use the same distance (14 Mpc) 
as M08. The measured disc scale-length of NGC 5907 is 6.1 kpc in $R$-band, while it is 3.82 kpc in $H$-band \citep{mr1995} 
when scaled to 14 Mpc. \citet{saha2009} got
3.86 kpc by fitting $Spitzer$ IRAC 4.5 $\mu$m data. \citet{barnaby1992}
obtained in $H$ band a scale-height about 0.41 kpc and a bulge-to-disc
luminosity ratio (B/D) of 0.05.

There are still considerable uncertainties in the stellar mass determination of NGC 5907. 
We need to re-estimate it carefully, accounting properly for the effects linked to the IMF 
\citep[see][]{puech2008,hammer2009}
Our stellar mass determination follows the method of \citet{bell2003} using the K-band observation from $2MASS$ and an optical colour, 
($B-V$) from \citet{just2006}. The $K_s$ apparent magnitude is
6.757, and the $B-V$ colour is 0.86. 

So the total stellar mass is 

\begin{equation}
log_{10} M_{\star}/M_{\odot} = -0.4 \times ( M_{K_s} - M_{K_s\odot}) + (-0.206) + 0.135 \times (B-V)
\end{equation} 

\noindent
where $M_{K_s\odot}$ is the solar absolute magnitude \citep{bell2003}. Then the total stellar mass of NGC 5907 is 6.57 $\times$ 10$^{10}$ M$_{\odot}$ assuming a $''$diet$''$  Salpeter IMF. 
The gas mass of NGC 5907 is about 1.94 $\times$ 10$^{10}$ M$_{\odot}$ including HI, H$_2$ and He \citep{just2006}. So the total 
baryonic mass is 8.51 $\times 10^{10}$ M$_{\odot}$, and the gas fraction is 23\%. 

The large scatter of values found in the literature are mainly due to
the different methods used to calculate the stellar mass. For example,
methods assuming that NGC 5907 lies just on the baryonic Tully Fisher relation
\citep[e.g.][]{mcgaugh2005,stark2009} are not sufficiently accurate for
an individual object. On the other hand, our estimate of the stellar
mass may underestimate the contribution of the bulge because NGC 5907
is seen edge-on and even at near-IR wavelengths, dust lying in the
disc may considerably affect its luminosity, and its precise contribution to the mass is still uncertain.

\section{Peculiar properties of the NGC 5907 halo}
\label{sec:haloproperty}
Even though NGC 5907 is a member of the 396th Lyon Group of Galaxies (LGG 396), all of the identified group members 
are at very large separation and could not be interacting with NGC 5907 \citep{im2006}. This galaxy may be 
considered as a typical isolated and an almost bulge-less spiral in the local universe.

However, recent observations show that the NGC 5907 halo has many peculiar features. 
\begin{itemize}
\item There are two giant loops lying near the polar 
disc plane of NGC 5907. \citet{shang1998} first discovered one half
of one loop, and more 
deep observations by M08 confirmed this result, revealing two giant loops. 
The two loops extend up to 50 kpc from the galaxy centre. This strange structure indicates that NGC 5907 has experienced 
a former merger/interacting event, which could be relatively recent according to M08. The surface brightness of these 
loops is about 26.8 mag/arc-second$^2$, corresponding to a stellar mass of approximately 3.5$\times10^8$ M$_{\odot}$ with 
a stellar mass surface density of 0.32 M$_{\odot}$ pc$^{-2}$.
\item The gaseous and the stellar discs are strongly warped. Previous observations have shown that the 
disc of NGC 5907 is sharply truncated at $\sim$ 24 kpc. The radius profile clearly shows a break, which is common 
to many local spirals \citep{van2007}. The deep observations of M08 reveal that the star light extends out to the nominal cut-off radius. 
The stellar disc is also warped \citep[see also][]{shang1998} in the same direction as the HI gas warp, suggesting
a common origin for both the stellar and gas warps. 
\item The halo of NGC 5907 is red according to \citet{lequeux1998}  
  with V-I=1.4 and B-V=1 at 5.6 kpc off the disc
  plane. According to \citet[see their Fig.9]{zibetti2004} this is
  slightly redder than the stacked value of about a thousand SDSS
  galaxies. This implies that NGC 5907 shares the red halo properties
  of many nearby spiral galaxies, including M31 \citep[see e.g.][]{mouhcine2005,hammer2007},
  indicating that its halo
  includes a significant fraction of metal enriched stars.
\item Mid-infrared observations revealed the presence of
  considerable material on both sides of the disc, up to 10 kpc from
  the galaxy centre, including PAH emission \citep{im2006} 
  molecular gas \citep{laine2010} and diffuse dust emission
  \citep{burgdorf2009}. This could indicate that the NGC 5907 disc
  is not completely relaxed.
\end{itemize}
Taken together, these abnormal features suggest that a former merger occurred in NGC 5907. Several 
models were set up to explain some of these features.

There are a few existing models for NGC 5907, which have set up to explain the formation 
process of the red halo or of the loops. To keep the disc undisturbed,
a minor merger with a high mass ratio has generally been proposed.  
 \citet{lequeux1998} model the red halo formation, assuming that it
 is from a merger with a red dwarf elliptical with mass of a few
 10$^9$ M$_{\odot}$.  \citet{rs2000} model the half
 ring discovered by \citet{shang1998} by using a model with a 1000:1 mass ratio.
A more recent model by M08 can reproduce both loop structures,
assuming that they formed through a single merger event with a mass
ratio of 4000:1. The small satellite, a spheroid, is assumed to fall
into the potential well of NGC 5907, forming a tidal tail that literally
draws out the two loops. Even though 
their model reproduces the geometry of the two loops well, it is
unclear whether it can reproduce all the abnormal features of NGC 5907
described above. It is also confronted with the problem of the
visibility of the progenitor residual.


We therefore investigate whether a gas-rich, major merger can
reproduce most of the abnormal features observed in NGC 5907. Such an
event may dramatically affect the whole structure of the main galaxy,
and its modelling should also reproduce the massive, inner
components (bulge, disc) of NGC 5907. To achieve this goal, we need to
consider a much larger number of constraints than that for a very
minor event. For comparison, we list in Table \ref{tab:comp}
the constraints and their possibly associated parameters in both major
and minor merger cases. It naturally leads to a huge parameter space
to be investigated for modelling NGC 5907 by a major merger, and a
natural difficulty of getting an accurate model. Our goal is to show
whether a single gas-rich merger event can reproduce the
global structure of the NGC 5907 bulge-less galaxy, 
as well as its exceptional system of loops in its halo. 

 \begin{table*}
 \caption{Comparing the number of constraints and parameters between minor and major mergers.}
 \begin{tabular}{lccc}
 \hline\hline
  Constraints    &  Major merger  & Minor merger & Model parameters	 	\\
 \hline
 Loops shape    &   Y            &  Y           & inclination, orbital parameters, view angle \\
 Loops mass     &   Y            &  Y           & mass ratio, gas fraction              \\
 B/T            &   Y            &  N           & mass ratio,star formation history, gas fraction\\
 Rotation Curve &   Y            &  N           & initial conditions		\\
 Gas warp       &   Y            &  N           & inclination, orbital parameters                   \\
Disk scalelength&   Y            &  N           & inclination,orbit parameters, mass ratio \\
 \hline
 \end{tabular}
 \label{tab:comp}
 \end{table*}

\section{Simulations and initial conditions}
\label{sec:simu}

We use the publicly available version of the GADGET-2 code \citep{springel2005}, which is a parallel TreeSPH code employing the fully
 conservative formulation \citep{sh2002} of smoothed
 particle hydrodynamics (SPH). In this code both energy and 
entropy are conserved even when the smoothing lengths evolve
adaptively \citep{sh2003}. Implementation of star formation, cooling, and feedback has been done following \citet{cox2006}, 
and our code is very similar to the one of \citet{cox2006} (see Appendix A).

We begin our simulations with a total baryon mass equal to 9 $\times 10^{10}$ M$_{\odot}$. 
This value is a few percent higher than the total 
baryonic mass of NGC 5907, to account for the mass loss during
the merger \citep{hammer2010}. The initial conditions 
are set up following \citet{hammer2010}. Each progenitor is
composed of a stellar and a gas disc embedded in a dark matter
halo. Both discs have an exponential distribution. The dark matter
model is chosen following \citet{barnes2002}.
The density profiles of these two components are  

\begin{equation}
    \rho_{\rm disc} \propto
      \exp(-R/R_{\rm disc}) \, {\rm sech}^2(z/z_{\rm disc}) \,, 
\end{equation}

\begin{equation}
        \rho_{\rm halo} \propto
      (r + a_{\rm halo})^{-4} .
\end{equation}

Following \citet{cox2006}, the gas-disc scalelength is set to three times 
that of stellar disc, since observations show that the gas disc has a 
more extended distribution than the stellar disc \citep{van2007}. 
The initial stellar disc scalelength is chosen following the scaling 
relation between rotation curve peak and disc scalelength \citep{hammer2007}. 
We notice that fitting the rotation curve leaves
a considerable degeneracy in the dark matter mass fraction in a galaxy
such as NGC 5907. Here we adopt nine percent of the baryonic matter, 
which is slightly more than other adopted values \citep{hoekstra2005,dutton2010} but less than those adopted in 
\citet{barnes2002}, \citet{cox2006}, and \citet{hammer2010}. The final scale
length of the dark matter profile is fixed using the baryonic Tully-Fisher relation 
 \citep{puech2010}. The disc scaleheight is set to be ten percent of the scalelength. 
We follow the method of \citet{barnes2002} to build the 
initial galaxy, which is stable enough for current work (see Appendix A). Parameters 
for each model are listed in Table~\ref{tabpar}.

 The adopted mass distribution of halo is more susceptible to tidal tail formation 
 \citep{dubinski1999,donghia2009}. To test the effect of the halo profile on the final result, 
we considered a Hernquist model \citep{hernquist1990}. This model is close to an NFW profile 
in the inner region \citep{sdh2005} and we find that the results do not change significantly 
(M3L34H model in Table \ref{tabpar}). The halo concentrations for the primary and secondary
 interlopers  are found to be C=11.48 and 15.6, respectively, the first value being very 
close to that of \citet{cox2006}, who model a gas-rich Sbc galaxy. We did not assume 
halo rotation since it is not expected \citep{sw1999} to have a strong influence on tail formation.
 Simulations use between 180000 and six million particles to test the effect of resolution. 

The gas fraction is assumed to be quite high (from 60 to 80\%), for progenitors that are 
intermediate mass galaxies at z$>$ 1. For comparison gas fractions of 60\% are found by 
\citet{daddi2010} for z$\sim$ 1.5 galaxies with stellar masses of 5$\times10^{10}M_{\odot}$ 
and similar values are found for massive galaxies at z$\sim$ 2 (Erb et al, 2006; see also Rodrigues et al. 2012 in preparation).  
To evaluate the mass ratio between the two merging galaxies, we follow the \citet{hopkins2010} 
zeroth order scaling relation between gas fraction, mass ratio, and bulge fraction. The 
small bulge component of NGC 5907 gives a constraint on the mass ratio,
which cannot be too small and should be larger than or equal to 2.

By construction, the present simulations assume a single encounter of
two galaxies, without accounting for other external supplies of gas
such as cold flows. Not being able to extract a galaxy similar to
NGC5907 from cosmological simulations has several consequences. For
example, the assumed gas fraction in the progenitors is certainly an upper limit, 
because some additional gas may be captured from the IGM before fusion.  The 
stability of the gas-rich progenitors when they are isolated has been verified in 
Appendix A, and do not account for gas accretion and clumpy fragmentation as  
could be inevitable with sufficient levels of gas accretion. Besides this, our simulations of an isolated disc reproduce those of Cox et al. for all configuration of density and  feedback. Possibly the absence of clump is related to ours and the Cox et al. recipes for SPH simulations; however, 
recent results from the AREPO code \citep{keres2011} show galaxies without numerous 
dense gaseous clumps, possibly contradicting with the numerous clumpy, intermediate-mass galaxies 
at z$\sim$ 2 \citep{elmegreen2009}. On the other hand, it is unclear whether clumpy disc progenitors would have changed our results after a major merger, which mostly destroy progenitor's features, except possible bulge-like structures that could be enhanced in clumpy galaxies  \citep{bournaud11}.

For some of our models, we assume a 
star formation history with a varying global efficiency in transforming gas to stars, 
in order to preserve enough gas from being consumed before fusion. Although this 
fine-tuned star formation history may have some physical motivations 
\citep[see e.g.][]{hammer2010}, its main role is also to ensure the formation 
of stars after the emergence of the gaseous disc just after fusion.

We adopt a retrograde-prograde merger, which is favourable for building
a disc after the merger \citep{hopkins2009}. The rotation direction
for the secondary progenitor is set to be prograde and that of the massive 
progenitor, retrograde. This choice is because resonances have a
strong effect on long tidal tail formation, which is enhanced in prograde and
suppressed in retrograde encounters \citep{tt1972,bh1992}. We also find that in order for the loop shapes
to match the observations, the orbital eccentricities should not be 
as low as found by M08, who used an orbital eccentricity of 0.42. 
We find, however that an orbit eccentricity of 0.9 is enough to have loop 
shapes that match the observations.

Cosmological simulations \citep{kb2006} provide a relationship
 between the ratio of the pericentre to the virial radius and the orbit 
eccentricity. We carefully choose pericentre and eccentricity so that our 
orbital geometry fulfils this relation. For example, we estimate the virial 
radius of the primary galaxy halo at the beginning of the simulation 
(8-9 Gyr ago, z$>$1) to be about 130 kpc from the relationship between 
virial radius and total mass derived from cosmological simulations \citep{sw1999}. 
Assuming a pericentre of 25 kpc implies eccentricity values of above 0.88 
\citep[see figure 6 of][]{kb2006}.

\section{Modelling NGC5907 and its loop system}
\label{sec:result}

\subsection{Formation of giant loops}
\subsubsection{Mechanisms of loop formation during the merger: setting the orbital parameters }

Appendix B describes the basic properties of loops formed in a major merger from 
tidal tail particles, which later on are captured by the gravitational potential 
of the galaxy.  Here we use these generic properties of tidal tails and loops to 
investigate different viewing angles and different forming epochs to reproduce 
the loops found by M08 in the NGC 5907 halo.

The assumption that the loops are well fitted by ellipses provides
important constraints on the viewing angles. To form long-lived loops,
their associated particles should not interact with the newly formed disc. We 
notice that one loop (noted as SW-SE-E1-W2 in M08) intersects the optical disc. 
Assuming it is a projection effect, this implies that the angle
between the line of sight and the loop plane should be 60 degrees or
larger. 
This is consistent with the M08 model, which assumed a loop rotation by
57 degrees. Moreover the inclination of the loop plane to the disc
plane is constrained 
by the observed tip position of the ellipse. The observed angle
between the tip of the SW-SE-E1-W2 and the direction perpendicular to
the disc, is quite small (less than 15 degrees), indicating that the
loop plane should be almost polar to the disc. The loop shape also
depends on the precession, intrinsic 
eccentricity, and position.

\begin{figure*}
\centering
\includegraphics[width=17cm]{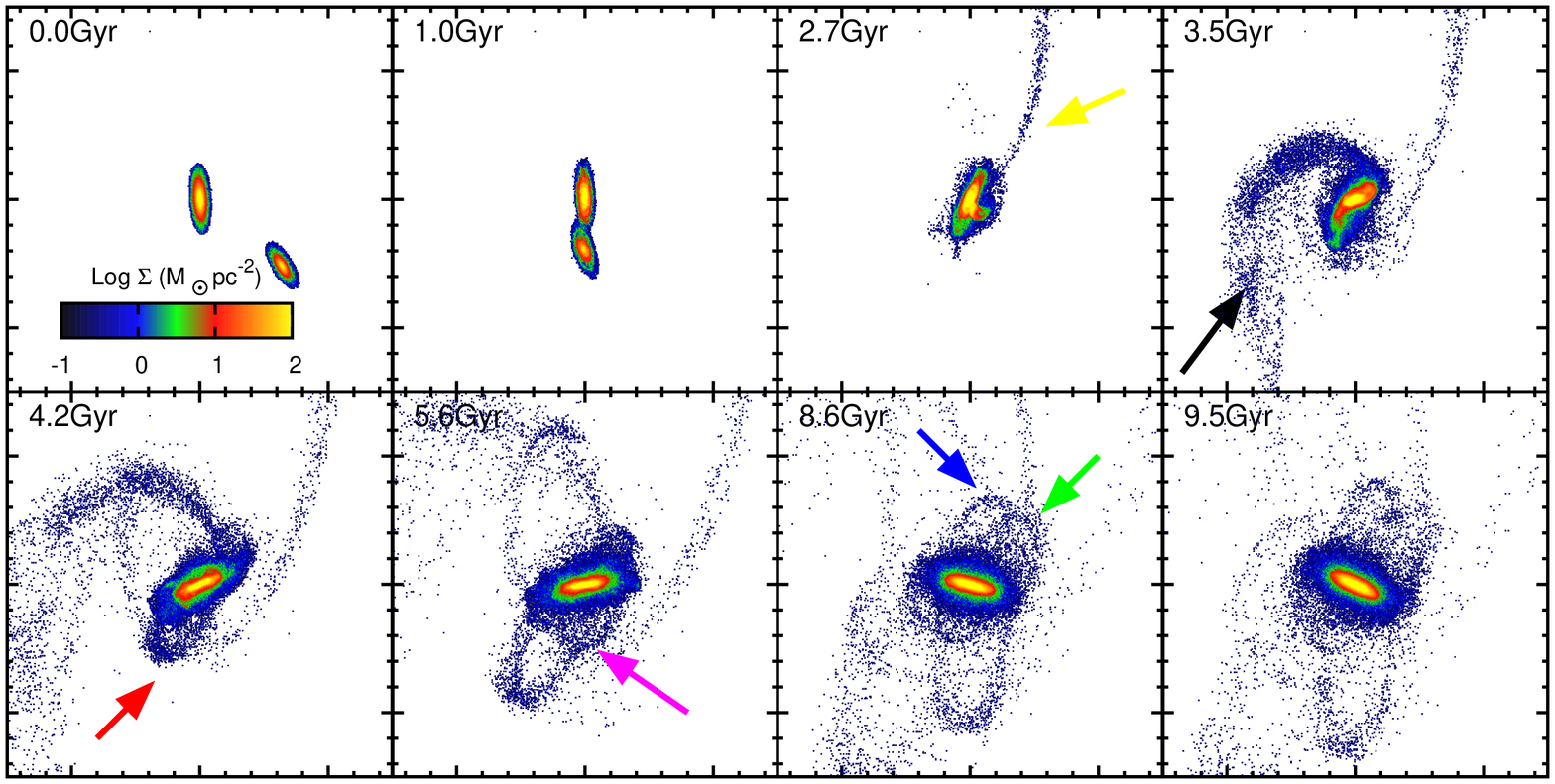}
\includegraphics[width=17cm]{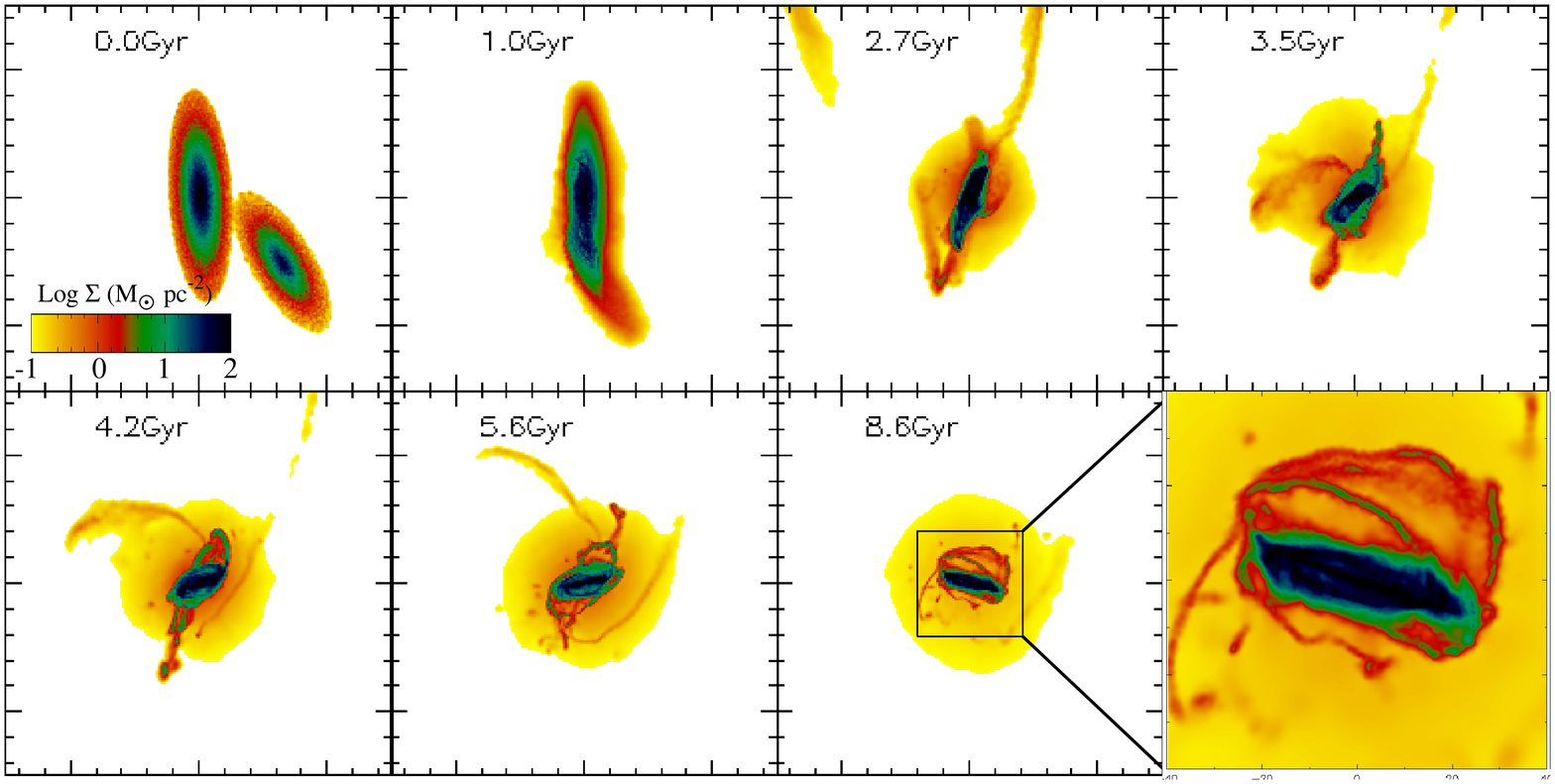}
\caption{Star and gas particles distributions for a merger with mass
  ration 3 (model M3L34A) at different epochs. Star particles are shown
  in the top two rows.  The fit to best observations is at 8.6 Gyr, at which
  the structures of third and fourth loops 
can match the observations. The gas surface density is shown in the
bottom two panels with gas density indicated by the colour bar. The 
size of each panel is 300 by 300 kpc and the arrows are explained in
the text.}
\label{fig:image3}
\end{figure*}

Table \ref{tabpar} summarises the properties of the models that are 
 used in this paper, including initial parameters, resolution, star
formation, or feedback history, as well as the structural parameters of the
remnant galaxy well after the merger. 
Feedback model 1 uses five times median feedback of \citet{cox2006} used. 
Feedback model 2 (M3L34F) uses five times this median feedback before fusion,and changes to a low feedback after fusion. 
The bulge Sersic index $n$ and disc scalelength are obtained by fitting the 
surface brightness density distribution with a Sersic profile, and two 
exponential functions. Model M3L34G6 changes gas fraction. Model M3L34H uses Hernquist 
model. Model M3L34A, M3L34B are similar to M3L34 after a fine tuning of galaxy inclinations. 
Figure \ref{fig:image3} gives an 
example of a 3:1 merger and displays its evolution 
at different epochs. The first passage occurs at
about 1.0 Gyr after the beginning of the simulation, and second  
at 2.7 Gyr. The first tidal tail is formed after the first passage, and expands at 2.7 Gyr, i.e. 
just before the second passage, as shown by a yellow arrow in the
upper third panel. We found that by increasing the pericentre radius, this tidal tail becomes 
less prominent and provides fewer stellar particles falling back to the newly formed galaxy. 
After the second passage, a second tidal tail is formed, as shown by a black arrow in the upper
right hand panel of Fig. \ref{fig:image3} (3.5 Gyr). Particles in this second tidal tail may 
form loops, and at $\sim$ 4.2 Gyr, the first loop begins to form (red arrow in the  
second left row panel). At this time the loop size is small. As time elapses an increasing 
number of high-elevation particles come back from the tail, and the first loop expands 
(see Appendix B). At $\sim$ 5.6 Gyr, the new disc is well formed, and a second loop 
appears (pink arrow in the 2nd row 2nd column panel). As time evolves, more and more loops
appear. At 8.6 Gyr, the third and fourth loops (green and blue arrows) are well formed,  
and their structures show some similarities with those of NGC5907. At the same time, the
first and the second loops have considerably expanded, which unavoidably leads to their dilution.

The two observed loops  have approximately 
the same apparent size, which we interpret to be a projection effect. Indeed, the higher 
order loop should be smaller than the lower order one, as described in Appendix B. A colour bar in the top left hand 
panel of Fig. \ref{fig:image3}
indicates the stellar-mass surface density. The surface mass density 
of the loops is about $\sim$ 0.1 to 1 M$_{\odot}$ pc$^{-2}$. M08 roughly
estimate the loops' surface brightness as $\sim 0.32 
$ M$_{\odot}$ pc$^{-2}$. Even though there is large uncertainty in the 
observed values, the model is able to reproduce the order of magnitude 
of the loop surface brightness.

\begin{table*}
\begin{center}
\caption{Eight models used in this study and their associated parameters. 
The parameters above the bar describe the progenitors, while those below it describe the merger remnant.
}
{\scriptsize
\begin{tabular}{llllllllllllll}
\hline \hline
parameters                  &M3L34G6&M3L34 &  M3L12&  M3L23&M3L34H&M3L34F&M3L34A&M3L34B \\
\hline                                                                                 
mass ratio                   & 3    & 3    &   3   &   3   & 3    & 3    & 3    & 3    \\
halo1 core size 	     & 11.8 & 11.8 & 11.8  & 11.8  & 20   & 11.8 & 11.8 & 11.8 \\
halo2 core size		     & 6.5  & 6.5  & 6.5   & 6.5   & 11   & 6.5  & 6.5  & 6.5  \\
stellar disc1 scalelength    & 4.6  & 4.6  & 4.6   & 4.6   & 4.6  & 4.6  & 4.6  & 4.6  \\
stellar disc2 scalelength    & 3.5  & 3.5  & 3.5   & 3.5   & 3.5  & 3.5  & 3.5  & 3.5  \\
Gal1 incx                    &-147  &-150  & -130  & -140  &-150  &-150  &-145  &-155  \\
Gal1 incz                    &-175  &-180  &  30   &  -70  &-180  &-180  &-170  &-180  \\
Gal2 incy                    & -30  & -30  &  -30  &  -30  & -30  & -30  & -30  & -30  \\
Gal2 incz                    &  0   &  0   &  0    &  0    &  0   &  0   &  0   &  0   \\
Gal1 gas fraction            & 0.6  & 0.8  &  0.8  &  0.8  & 0.8  & 0.8  & 0.8  & 0.8  \\
Gal2 gas fraction            & 0.8  & 0.8  &  0.8  &  0.8  & 0.8  & 0.8  & 0.8  & 0.8  \\
$r_{peri}$                   & 25   & 25   &  25   &  25   & 25   & 25   & 25   & 25   \\
eccentricity                 &0.9   &0.9   & 0.9   & 0.9   &0.9   &0.9   &0.9   &0.9   \\
Feedback                     &  1   &  1   &  1    &  1    &  1   &  2   &  1   &  1   \\
N$_{particle}$               &2.21M & 2.21M& 180k  & 630k  & 630k & 1.26M& 2.21M& 1.26M \\
m$_{dm}$:m$_{star}$:m$_{gas}$& 4:1:1& 4:1:1& 4:1:1 & 4:1:1 & 4:1:1& 4:1:1& 4:1:1& 4:1:1\\
\hline                                                                                 
Observed time (Gyr)          & 8.0  & 8.6  & 5.6   & 7.3   & 8.6  & 8.6  & 8.6  & 8.6  \\ 
Disk scalelength(kpc)        & 4.38 & 4.52 & 3.58  & 3.93  & 4.22 & 5.24 & 4.76 & 4.36  \\
Bulge sersic index($n$)      & 1.2  & 1.4  & 0.9   & 1.0   & 1.6  & 1.7  & 1.4  & 1.4  \\
Re of bulge (kpc)            & 0.98 & 1.22 & 1.05  & 1.32  & 0.97 & 0.37 & 1.19 & 1.11 \\
B/T (fitted)		     & 19\% & 23\% & 14\%  & 19\%  & 23\% & 15\% & 25\% & 25\% \\
Final gas fraction	     & 32\% & 41\% & 53\%  & 46\%  & 43\% & 16\% & 42\% & 43\% \\
\hline                       
\label{tabpar}               
\end{tabular}                
}                            
\end{center}                 
\end{table*}

\begin{figure}
\centering
\includegraphics[width=9cm]{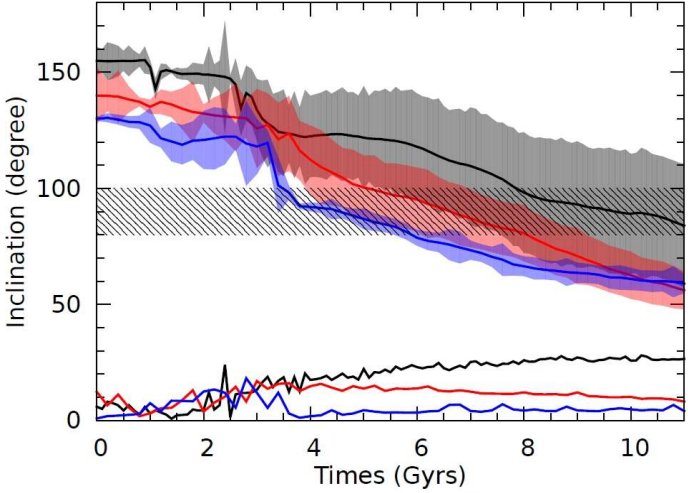}
\caption{Inclination angle between disc, loop, and orbital planes as a
  function of time. The lower 
solid lines show the angle between the loop and the orbital planes. Lines within 
shaded regions indicate the inclination angles of the disc plane relative to 
the orbital plane, while the line-shaded region around 90 degrees indicates the 
angles between disc and loop planes that are consistent with the observations. 
Different colour lines show the simulations with different initial disc inclinations 
to the orbital plane. All the models used here have a mass 
ratio of 3 (from bottom to top, models M3L12, M3L23, M3L34B).}
\label{fig:diskplane}
\end{figure}

Figure \ref{fig:diskplane} shows how the disc and loop inclinations
evolve 
with time. The loops are formed by particles coming back from the
tidal tail. We quantified the loop plane inclination relative 
to the orbital plane from the angular momentum of the tidal tail
particles. Because the angular momentum of the tidal tails is
dominated by the orbital angular momentum, tidal tail particles 
mostly lie on the orbital plane, and at most 20 degrees away from 
it. To quantify the disc plane, we use the angular momentum of the 
post-fusion stellar particles, following the procedure presented in 
\citet{hammer2010}.  With evolving time, the disc plane systematically
rotates so as to be aligned to the orbital plane. Using the 
inclinations of disc and tidal tail particles relative to the 
orbital plane, we can determine the inclination between the loops 
and the disc plane that is consistent with observations. The angle 
between loop and disc planes is about 80 degrees in M08, while 
 \citet{rs2000} use an angle of 90 degrees to fit 
one loop. We considered this ten degree difference as the uncertainty 
in the angle between the loop and disc planes, as well as the uncertainty 
in our models, so in Fig. \ref{fig:diskplane} the acceptable angles 
are shown by the line-shaded region. Because higher order loops 
appear at later epochs, the initial disc inclination 
angle should be larger to make the disc and loop plane close to polar at later times.

\begin{figure*}
\centering
\includegraphics[width=15cm]{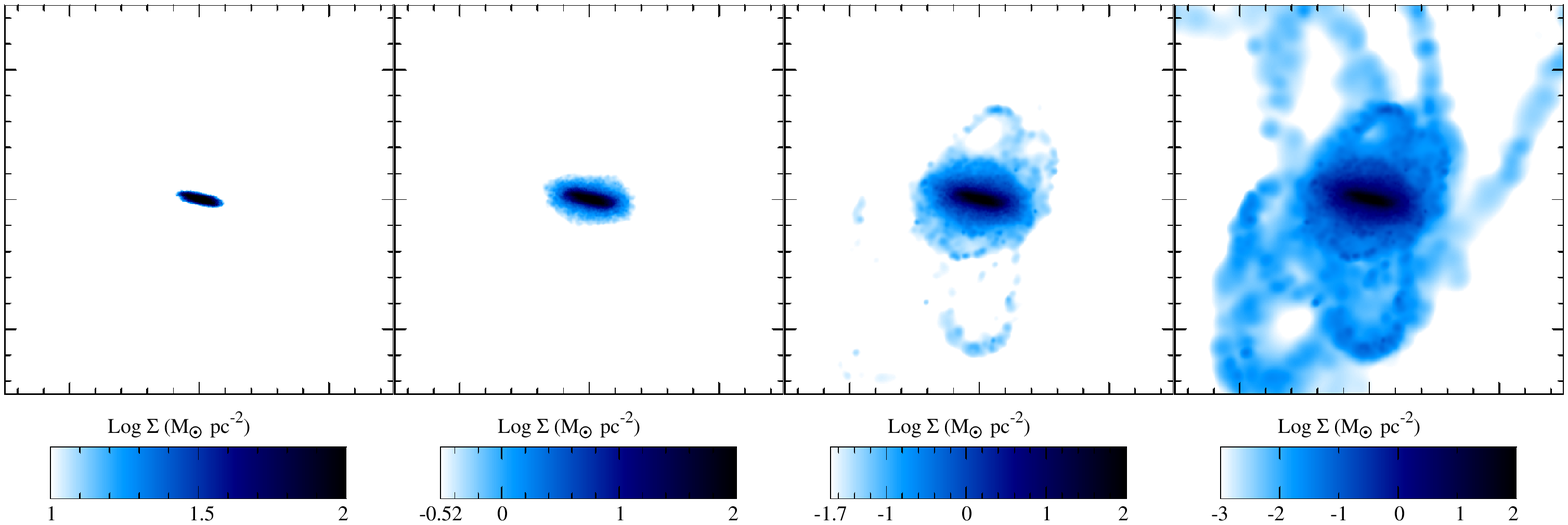}
\caption{Stellar mass surface density shown with different scales of stellar 
mass surface brightness at 8.6 Gyr for the model M3L34. Each panel size is 300 by 300 kpc.}
\label{fig:threshold}
\end{figure*}


\subsubsection{Comparison with observations: which loops matter ?}

As shown in the previous section, a major merger can generate loops
with a geometry similar to the observed ones, at a time that 
depends on the initial orientation of the main progenitor disc
plane. Because observations are surface brightness limited, we
illustrate in Fig. \ref{fig:threshold} how the visual impression 
depends on the surface-brightness threshold. By varying the threshold, the thin
disc, thick disc, and loop features are progressively revealed. Colour
bars in Fig. \ref{fig:threshold} indicate that loop regions with surface brightness 
ranging from 0.1 to 1 M$_{\odot}$ pc$^{-2}$ coincide quite well with
 the observed loops. Since the loop surface brightness is 0.32 M$_{\odot}$ pc$^{-2}$, 
this argues that observations identify only the brightest part
 of the loop system, i.e. mostly above the galaxy in the panel
 corresponding to 8.6 Gyr. 

After the merger, particles coming back from the tidal tail  
continuously form loops of increasing order as illustrated in
Fig. \ref{fig:image3}. During this process, while
higher order loops are forming and reach sizes comparable to the
observed loops, lower order loops continuously expand and become more
and more diluted with time. Another factor favouring the low-order
loop dilution is related to the tidal tail surface brightness
distribution, which is fainter farther from the galaxy. The low-order
loops are formed at any time by particles coming back later, which results
in lower surface brightness compared to high-order loops.

\begin{figure*}
\centering
\includegraphics[width=17cm]{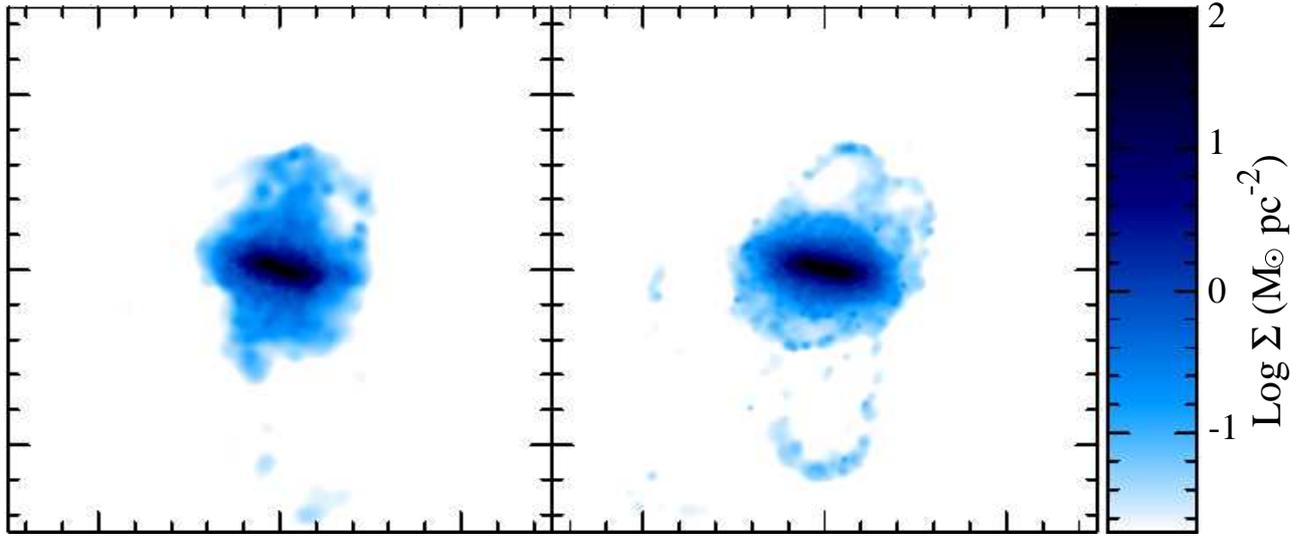}
\caption{Comparing loop structures formed by combining loops of
  different order for mass ratios of 3. The left panel shows the 
first and second loops (model M3L23), the second panel shows 
the third and fourth loops (model M3L34), while the first loop 
appears at the bottom of the galaxy and becomes diluted. The size 
of each panel is 300 by 300 kpc. The loop stellar surface mass 
densities are consistent with observations from M08. In all these 
images the threshold has been chosen slightly below the observed 
one from M08.}

\label{fig:imgcomp}
\end{figure*}

We also find that a more inclined initial disc can capture the
material of the secondary interloper better by making it rotate closer to the disc plane. 
Thus, there is little residual signature when the disc is seen in an edge-on position
and favours observed loops of a high-order type. This is 
illustrated well in Fig. \ref{fig:imgcomp}, which compares the formation of
different order loops at different epochs. Residuals from the secondary
decrease dramatically with time, favouring high-order loops that match
the observations. Figure \ref{fig:imgcomp} also illustrates that lower
order loops almost vanish with time.

In our model, the left loop has a higher order than the right one. 
This is in good agreement with observations, since the left loop 
has been discovered first by \citet{shang1998}, and the discovery of the two loops by M08 has
been done with much deeper observations, implying that the left loop is
brighter than the right one and that surface brightness threshold has
to be accounted for when modelling the loops.

Another strong argument favouring observed loops of the high-order 
kind is related to the nature of the progenitors, which are
assumed to be gas-rich galaxies to allow the formation of a
significant disc. Such gas-rich galaxies are common in the
high-redshift universe,
and far less common at intermediate redshifts.  Using this
argument, together with the absence of residuals, favours a
scenario for which the observed loops coincide with a combination of
at least third and fourth order loops, with a time after interaction
longer than $\sim$ 8.6 Gyr. Interestingly, this scenario can be falsified
by further observations: deeper observations of a wider field of view
should detect the fainter lower order loops.

\subsection{The rebuilt central galaxy: disc and bulge}

\begin{figure*}
\centering
\includegraphics[width=16cm]{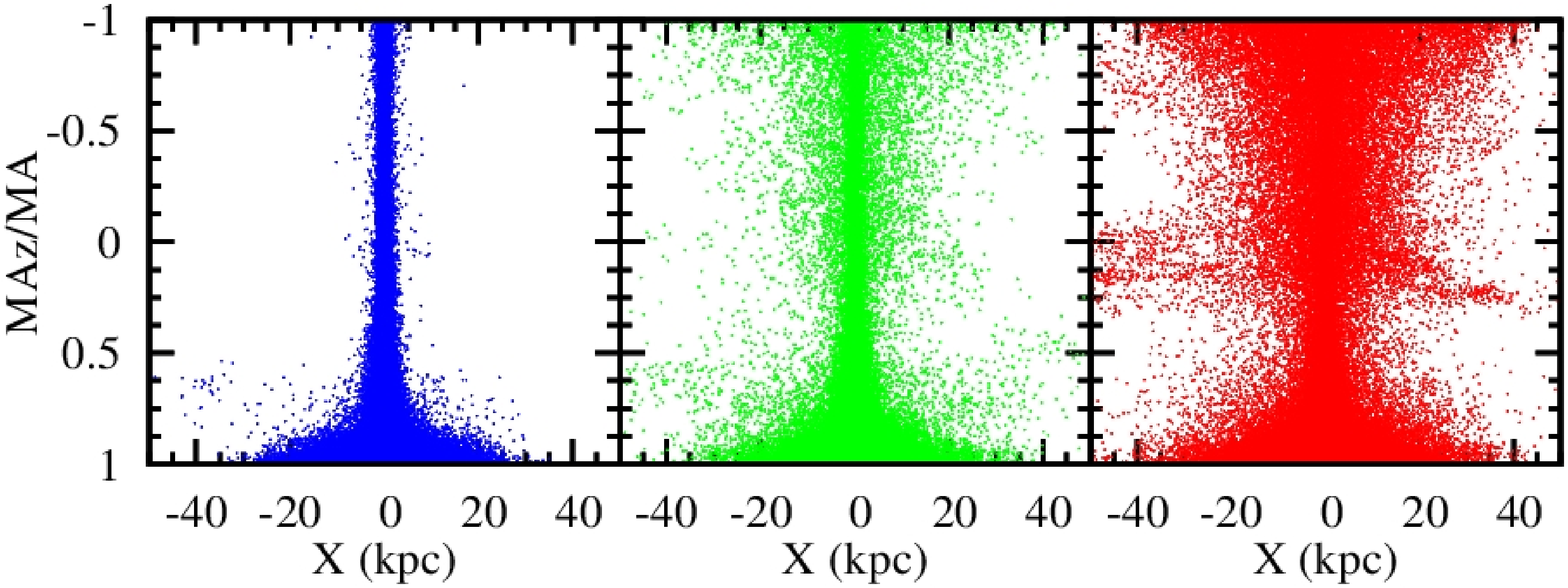}
\includegraphics[width=10cm]{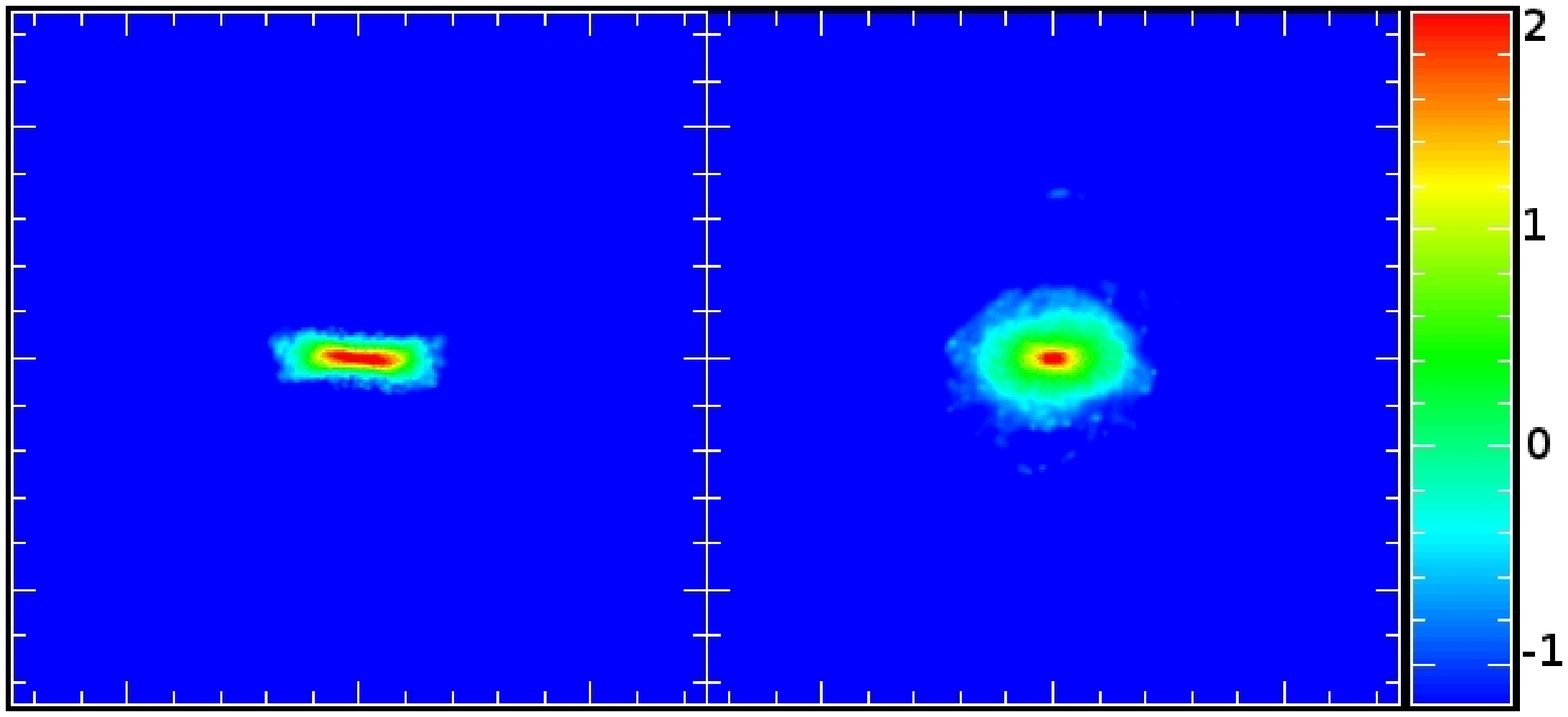}
\caption{The top panel gives the distribution of $z$ component of the 
angular momentum of stellar particles for model M3L34. The galaxy has been 
projected edge-on using the angular momentum of the baryons. From left to right, 
star of different age (from young, intermediate, and old) are shown. Old stars (red dots) already exist
in the progenitors at the beginning of the simulation. Young stars
(blue dots) are defined with ages over 3 Gyr after the
beginning of the simulation. Ages between 
young and old are defined as intermediate stars (green dots), which are formed before the fusion. The
bottom panel shows the projected mass distribution of the thin disc
(left), and the bulge and thick disc (right). The thin disc is selected 
from its angular momentum ratio (MAz/MA$\ge$ 0.9, see Hammer et al, 2010),
 which cuts the warp off.  Each panel has a size of 300 by 300 kpc.}
\label{fig:decomp}
\end{figure*}

To test whether the final merger remnant is consistent with the
disc of NGC 5907, we follow the decomposition method of \citet{hammer2010}. 
Each final remnant is decomposed into three-components, 
namely bulge, thick disc, and thin disc. This method is based on 
angular momentum distribution taking advantage of the full three 
dimensional information for each particle \citep[see details in][]{hammer2010}. 
Figure \ref{fig:decomp} gives an example for a 3:1
merger model. The top panels show the angular momentum distribution 
of young, intermediate age, and old stars, which are defined following \citet{hammer2010}. 
Most young and intermediate age stars have angular momentum along the disc polar axis 
(0.9$\le$MAz/MA $\le$1) or the thick 
disc (0.7$\le$MAz/MA $\le$0.9), with the rest of them mostly
concentrated in the bulge. The top right hand panel shows the presence of
many old stars at a large distance from the centre, with angular
momentum very different from that of the disc. They are mostly
particles of the loops, but not do rotate with
the disc. The bottom panels of Fig. \ref{fig:decomp} show the projected
 mass distribution for different components, confirming the presence of 
a prominent thin disc including $>$ 70\% of the baryonic mass after the 
merger event (left panel).

We also fitted the stellar surface brightness distribution similarly
to what is usually done from observations.  For each merger remnant, 
 we fit the galaxy surface brightness with a bulge that has a free Sersic 
index and an exponential disc (Sersic n=1). For each merger remnant, 
we define the disc plane using the three-dimensional information as 
defined after the galaxy decomposition using the angular momentum 
(see Fig. \ref{fig:decomp}). The stellar surface density is then 
measured in logarithmically spaced annuli in the disc plane.

\begin{figure*}
\centering
\includegraphics[width=7cm]{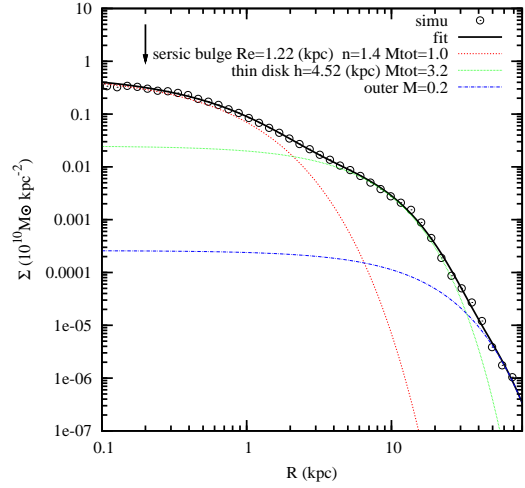}
\caption{Stellar surface mass density distribution as a function of radius for model M3L34. Open circles show 
the total stellar density together with a fit by different components (solid line). Sersic bulge and 
exponential disc components (indicated by different colours) are used to fit the stellar mass surface
 brightness. The fitting process excludes an area with a radius equal to 2 times the softening 
length (labelled by arrows, see the text). }
\label{fig:diskh}
\end{figure*}

Figure. \ref{fig:diskh} shows one example of a fit for a model with a mass ratio of three. 
During the fit, we excluded the very central part because it could be affected by numerical softening.
 We note that an additional exponential component is necessary to fit the 
outskirts. This additional component presents a surface brightness similar to
or fainter than that of the loops and corresponds to the inner halo of the simulated galaxy.
The results of the different models are listed in Table \ref{tabpar}. The disc scalelength of our models 
ranges from 3.9 to 4.5 kpc and is consistent with the observed scalelength value of 3.84 kpc. 
The bulge fraction ranges from 15 to 23\%, and depends on both the model and the
star formation history. We caution that this bulge fraction is derived from the stellar 
mass distribution, while in some models $\sim$ 40\% of baryons are still made of gas.

Observations show that NGC 5907 is relatively gas-rich  (gas fraction
of 23\%), which can be compared to the final gas fractions of the different models 
listed in Table \ref{tabpar}. Models with constant feedback result in excessive 
gas fraction in the final remnants. We verified that the final gas fraction may 
reach the observed value using a tuning of the feedback (or an increase in the 
star formation efficiency) just after the fusion (model M3L34F), as assumed by \citet{hammer2010}. 
This also leads to a lower bulge fraction, with B/T=15\%.

The surface-brightness fitting method also gives the bulge half light
radius and the bulge sersic index. Most of our realizations of gas rich
major mergers result in low Sersic index values ($\lesssim$ 1.5),
i.e. comparable to those of pseudo bulges \citep[e.g.,][]{kk2004}.



\subsection{Rotation curve}
\label{sec:rc}

\begin{figure}
\centering
\includegraphics[width=8cm]{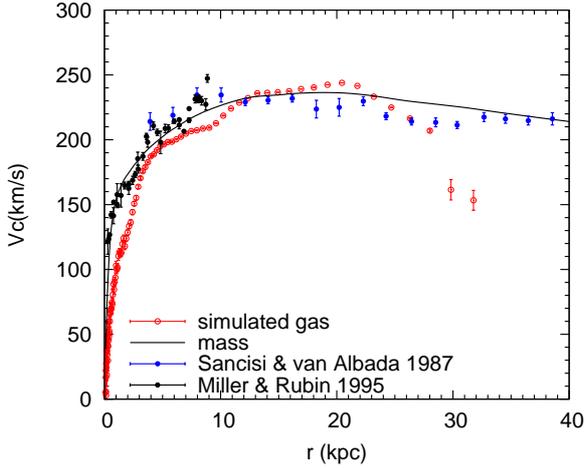}
\caption{Comparison of observed and simulated (model M3L34A) rotation curves. The blue points are 
from \citet{sv1987} and are based on HI observations. Black solid points are from $H\alpha$ and 
NII observations by \citet{mr1995}, including near the central regions. Solid line shows the prediction from the 
mass model V$_c=\sqrt{\frac{GM(<r)}{r}}$, red points are direct measurements from the gas kinematics 
within $\pm$1.5 kpc from the disc plane.
} 
\label{fig:rc}
\end{figure}

We also compared the model rotation curve with observations. We collected two observations
 for rotation curve of this galaxy as shown in Fig. \ref{fig:rc}. 
Even though the observations show some discrepancies between them, 
our model reproduces the observed rotation curve quite well. 

\subsection{ Disk warp and disturbed material above the disc plane}

The merger remnant is not very relaxed at the time of observation,
which makes the disc dynamically hot. This violent process 
naturally results in a disc warp and disturbances. The bottom rows of Fig. \ref{fig:image3} show the gas 
evolution after the merger and the presence of a warp at 8.6
Gyr. After the merger, gas is decoupled from stars owing to their
different dissipational properties, which is consistent with the
observational result that the loops only include stars. 
The merger process makes the gas warped at the edge of the disc, and this can be
maintained for several Gyr. \citet{shang1998} did a 6 hr radio
observation with the VLA. In their figure 2 of an 
integrated 21 cm intensity map, the gas distribution is very warped
and extends above and below the plane of the disc.

The gas above the disc plane shows non-relaxed motions that are also expected with our 
modelling of a major merger. \citet{im2006} analysed the 
PAH distribution and found PAH emissions well above the disc, up to
8.3 kpc. CO observation of the central parts of the disc 
show a steeply rising rotation curve and non-circular 
 molecular gas motions, presumably due to a bar \citep{garcia1997}. These observations could be consistent with a 
major merger origin of NGC 5907.

\subsection{Effects related to resolution}

To test the effect of resolution, we compared simulations with different
numbers of particles, from 180 k particles to six million for model M3L23. 
This test is used to verify whether weighing dark matter particles and gas
particles affects our final result differently. In this study, 
the mass of the individual particles for dark matter, gas, and stars
has ratios from 10:1:2 to 4:1:1. By varying particles mass ratio and increasing 
total particles number up to six million, we find that our
results, including the structure of the loops, are not significantly
 affected by resolution.

\section{Discussion and conclusions}
\label{sec:discuss}

\begin{table*}
\caption{Comparing major and minor merger hypotheses, where "Y" means consistency with the observations.}
\begin{tabular}{lcccl}
\hline\hline
Observational features    		  &  Major & Intermediate & Minor & Comments                                              \\
 & merger & merger & merger & \\
                & 3:1-5:1 & 5:1-12:1 & $>$ 12:1 & \\
\hline
Loop shape    		  &   Y?  & Y?          &  Y           & In minor merger, the loops trace the progenitor orbit, while in a  \\
                          &                &       &       & major merger, loops are formed by particles coming back from a tidal tail        \\
Loop size                 &   Y     & Y      &  Y           & \\
Loop surface mass density &   Y     & Y       &  Y           & Most models can reproduce the stellar mass surface density \\
Loop eccentricity & Y & N? & N & With an orbital eccentricity of 0.9-1, a  merger with mass ratio $>$ 12:1 \\
 & & & & would need more than a Hubble time to reach fusion\\
Visibility of             &   Y      & Y      &  N           & Major mergers provide a thick disc component after complete fusion of the nuclei, \\
  remnants                &        &        &              & and low dynamic friction means the remnant nucleus should be seen in a minor merger \\
Gas warp       		  &   Y     & Y       &  N           & Major merger predicts a gas warp unlike a minor merger with high mass ratio \\
Gas \& PAH emission       &   Y   & Y ?        &  N           & Major merger predicts residual gas and PAH emission above the disc, and \\
above the disc            &          &      &              & this should be explained by another mechanism in a minor merger\\
Colors of loops           &   Y      & Y      &  N?          & For a minor merger the satellite should be a relatively massive, red dwarf elliptical\\
\hline
\end{tabular}
\label{tab:majorminor}
\end{table*}

\subsection{Can the properties of NGC5907 be reproduced by an 
ancient and gas-rich major merger?}

We modelled NGC 5907 as the result of a gas-rich major merger, using
progenitors with a mass ratio of three. With gas fractions in the progenitors
ranging from 60 to 80\% and a star formation history similar to the one in
\citet{hammer2010}, our modelling is able to reproduce the main observed
NGC 5907 structural parameters (the thin disc and its rotation curve,
and the bulge). The exceptional features of NGC 5907 can be
reproduced, together with the central galaxy properties, especially if
we compare the observed loops to the high-order loops
expected in a major merger model. Given the extremely large number of
parameters, as well as the very numerous constraints provided by the
observations, we cannot claim that we have already identified the
exact and unique model of NGC 5907 and its halo properties. We
nevertheless succeeded in reproducing the loop geometry, and  
a disc-dominated, almost bulge-less galaxy. The limitations of our 
modelling are two-fold, with one related to observational uncertainties, 
the other to our modelling procedure.

\begin{figure}
\centering
\includegraphics[width=7cm]{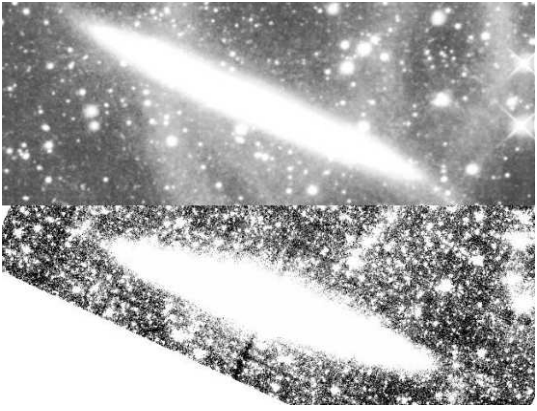}
\caption{{Top:} Deep observations of NGC 5907 in the visible light
showing the thin disc with the gigantic loops (from M08). {Bottom:} Observations of 
NGC5907 with IRAC at 3.6$\mu$m revealing the whole extent of the disc that is 
twice thicker than in the visible light. 
}
\label{fig:OptIR}
\end{figure}

Distance uncertainties of $\sim$ 30\% translate into uncertainties in
the total baryonic mass of the galaxy, which may affect the accuracy
of our modelling. However, the main uncertainty when comparing observations to modelling is
related to the striking difference of the NGC 5907 disc at visible and near-IR
wavelengths as illustrated in Fig. \ref{fig:OptIR}.  At visible
wavelengths the
thin disc axis ratio is almost eight, the bulge shows a negligible
contribution, while this exposure is deep enough to detect the very
low surface brightness loops. This is in sharp contrast to 
observations at near-IR wavelengths, which reveal a thicker disc, with
an axis ratio of 4.2, and do not detect the loops.

Two effects can explain the above behaviour of the edge-on disc of NGC 5907. Firstly,
it has been shown that thick discs are prominent at near-IR wavelengths \citep[see e.g.][]{comeron2011}, 
and secondly, the presence of diffuse interstellar dust up to 10 kpc above the disc 
has been identified in NGC 5907 \citep{burgdorf2009}. At 5 kpc from the disc, we 
estimate the surface brightness of the disc to be 22.5 $mag$ $arcsec^{-2}$ at 
3.6$\mu$m, which translates into 25.3 $mag$ $arcsec^{-2}$ in the AB system. At 
this location, there is a very faint emission in the image by M08 with $\sim$ 28.0 $mag$ $arcsec^{-2}$ in the R band, i.e.  
28.3 $mag$ $arcsec^{-2}$ in the AB system. Such a huge gap of three magnitudes between 
near-IR and visible light is probably caused by a combination of both effects, very red 
stars in the extended thick disc, and dust attenuation in the visible. It translates 
to $f_{\nu}$(V) $\sim$ 0.06$\times$$f_{\nu}$(3.6$\mu$m). To estimate the impact of 
dust extinction at 5 kpc on both sides of the disc, we  assume that the thick disc stars 
cannot be intrinsically redder than M stars. For a non-extincted M star with $T_{e}$=3500K, 
\citet[see their Fig. 7]{benjamin2007} find $f_{\nu}$(V) $\sim$ 0.15$\times$$f_{\nu}$(3.6$\mu$m). 
Thus even if the thick disc was only populated by M stars, it should have been detected 
in the very deep exposure of M08, except if the visible light has been extincted by a 
factor 0.15/0.06=2.5, which corresponds to $A_{V}$=0.9 (Galactic extinction law for R=3.1). For illustration, by replacing M by K stars (with $T_{e}$=4500K) would lead to an extinction factor of 12 instead of 2.5, emphasising the importance of dust extinction on both side of the edge-on disc.
Ideally, only a very deep exposure at near-IR wavelength can provide an image
that can be compared to the stellar mass distribution of a model.

Another substantial uncertainty from observations is related to the
B/T estimate. Our model predicts B/T values around 20\%, and a
specific implementation of varying star formation history (or varying
feedback history) reduces this value to 15\%. The latter
value is approximately twice the value we derive in fitting the
luminosity profile of the 3.6 $\mu$m image (B/T $\sim$ 0.08), which
should be less affected by dust than at visible wavelengths. In fact
our simulated B/T for the stellar mass may not coincide exactly with
estimates from the luminosity. It has been argued, for example, that the
IMF in elliptical galaxies is more dominated by low-mass stars than
that of the Milky Way disc \citep{vc2010}. Given the
similarities between ellipticals and bulges, one may speculate that the
NGC 5907 bulge includes a larger fraction of low mass stars than the
disc, leading to a B/T similar to the simulated one.

\begin{figure}
\centering
\includegraphics[width=9cm]{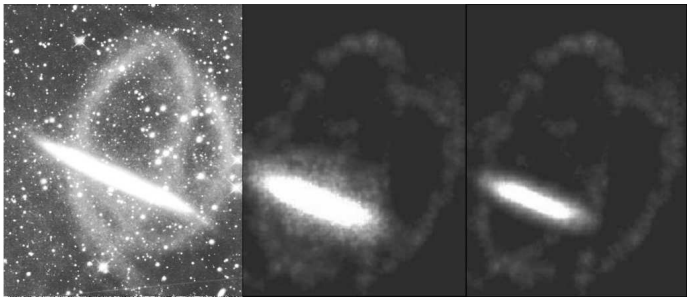}
\caption{{Left:} Deep observations of NGC 5907 in the visible light
showing the thin disc with the gigantic loops (from M08). {Middle:} 
One simulation of the loops at T=8.0 Gyrs with the M3L34G6 model, in 
which we assume a dust extinction screen affecting the stellar light 
density by a factor of 2.5 on both sides of the edge-on disc 
(see discussion in Sect. 6.1). {Right:} Same model as middle panel 
shows all components with accounting for an extinction by a factor 12.
The real extinction should be between the middle and right panel.}
\label{fig:disks-loops}
\end{figure}

Limitations of our modelling are linked to the huge parameter space 
to be investigated. It leads to loops that 
behave like the observed ones, with some differences in the size
and geometry. In Fig. \ref{fig:disks-loops} the modelled loops also
show some differences with the observed ones when comparing their
locations with respect to the disc. This indicates that, while we
correctly identify the mechanism for loop formation, we do not yet recover
  the best solution for NGC 5907, which would require an enormous 
amount of time for fine tuning the models. Other limitations come from 
 (1) the use of GADGET-2, which, for given resolution and star formation, 
feedback, and cooling recipes, provides less well defined discs
\citep[see Fig. 11 of][]{keres2011} with lower angular momentum than
those predicted by the AREPO code.  (2) The fact that our simulations 
are done in isolation, while other sources of gas supplies may be brought 
to the galaxy as done in cosmological simulations. Accounting for the 
above could have led to more realistic discs, and as it reduces 
the need for too high a gas fraction in the progenitors.

\subsection{Comparing the relative merits of minor and major merger models}

As mentioned above, the main drawback of the major merger scenario is
caused by the enormous number of constraints to account for, leading to
simulated configurations that still show some discrepancy with
observations. This is in sharp contrast to a minor merger case for
which the satellite orbit is only constrained by tracing the loops,
without any influence on the main galaxy. This point, however, is mostly
technical and should be discarded when discussing the relative
relevance of both scenarios.

Table \ref{tab:majorminor} summarises the exceptional features
revealed by deep observations of NGC 5907 at visible, mid-IR, and radio
wavelengths, and displays the merits of major, intermediate, and minor
mergers, respectively, to reproduce them. To sample a wider range of
mass ratios, we ran simulations of 10:1 and 15:1 mergers assuming
a pericentre radius of 25 kpc and orbits consistent with the formation
of loops similar to the observed ones. While the 10:1 merger occurs at
T= 9 Gyr after the beginning of the simulation, the 15:1 merger would
take more than a Hubble time to effectively merge. Thus we estimate
that 12:1 is a limit for reproducing NGC 5907 with the orbital
parameters expected from cosmological simulations.

A possible advantage of the major merger is linked to 
 its being likely to predict all these features together,
while the minor merger case can only reproduce the loops. Features
like the warp or the gas and PAH emission above the disc can be 
explained well by other mechanisms, such as additional interactions or
ejection of dust particles from the thin disc by galactic winds.
 The presence of considerable amounts of non-relaxed
material (dust, gas including in molecular phase) in the disc
outskirts, as well as the strong warp favours a scenario for which the
NGC 5907 is not a relaxed disc, but this cannot lead to a decisive conclusion.

Possibly the absence of a remnant signature, together with loop colour are
more compelling for distinguing between the two scenarios. Using the IRAC
3.6$\mu$m image \citep{ashby04}, we were able to model the light distribution with
GALFIT \citep{peng2010} using a bulge, plus disc
decomposition. After subtracting the model from the image, we 
examined the presence of a possible residual of the progenitor  of a
minor merger at the locations where the loops intersect the edge of
the disc. At such a large distance (20 kpc) from the disc centre, it is
unlikely that dust can affect the light that much, so we find no object
that could be responsible of such an event. This is really problematic
for a minor merger scenario, because with such an extended orbit, it is
unlikely that tidal forces could destroy it. In the major merger case
we have shown that the low surface-brightness features left by the
secondary interloper gradually vanish with evolving time, because they are 
progressively captured by the disc (see Fig. \ref{fig:imgcomp}).

\begin{figure}
\centering
\includegraphics[width=8cm]{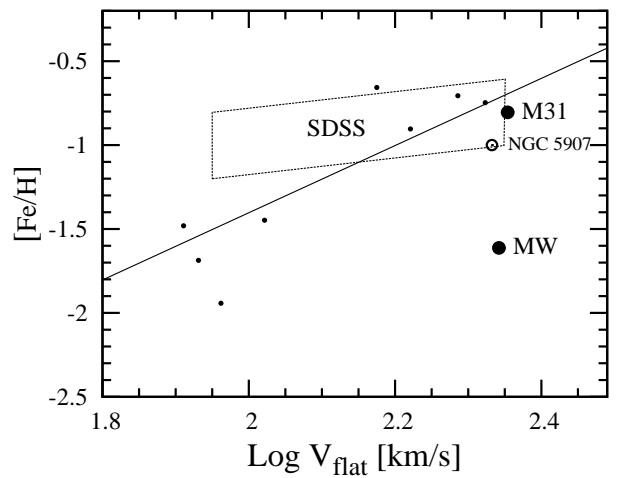}
\caption{Comparing iron abundances of outskirts of NGC 5907 with that of other galaxies
 from \citet{hammer2007}. The iron abundances of the NGC 5907 inner halo are from \citet{zheng1999},  
who estimate them by comparing the loops colour with that of Galactic globular clusters. }
\label{fig:iron}
\end{figure}

The colour of the loops has been estimated by different groups and
 confirmed to be both red \citep{zheng1999,lequeux1998} and
 comparable to that of elliptical galaxies \citep{zibetti2004}. With
 the red colour $R-I\sim 0.5\pm 0.3$,  \citet{zheng1999} found it is
 consistent with Galactic globular clusters with [Fe/H] $\sim
 -1$. Following \citet{hammer2007}, we show  in Fig. \ref{fig:iron} the relation between 
 rotational velocity and colour-inferred iron abundances of the stellar disc outskirts.
NGC 5907 follows the same relation as defined by most spirals, which
could be easily explained if stars in the inner halo are enriched by an ancient
major merger \citep[see][]{hammer2007}. Given their location in the mass 
metallicity relation, a dwarf spheroid origin for the loop progenitor would generate a bluer colour 
 than what is observed. For example, the colour of Sagittarius and its 
stream [Fe/H]=-1.2 \citep{sesar2011} hardly matches the colour of the NGC5907 halo.

Assuming a cosmological origin for the encounter that is responsible
for the loops has important consequences. In fact, this implies orbit
eccentricities in excess of e=0.85 \citep[e.g.][]{kb2006} for a pericentre 
larger than 25 kpc. This would exclude the M08 model that assumes e=0.42, 
as well as all models with mass ratio over 12:1, simply because 
the two galaxies would not have time to merge within a Hubble time, leading 
 the problem of the absence of progenitor residual too critical. 

Examination of Table \ref{tab:majorminor} confirms  the above: only
mergers with mass ratios between 3:1 and 12:1 are consistent with most
of the observations of the NGC5907 halo. Given the observational
constraints and our modelling, it is not possible to distinguish
between the merits of intermediate mergers (5:1 to 12:1) and major
mergers (3:1 to 5:1). In fact, we have failed to reproduce the observed loops 
with a 10:1 merger, although we certainly did not investigate the whole 
range of parameters to exclude this. On the other hand, with orbital 
eccentricities from 0.9 to 1, less dynamical friction is expected for a 
higher mass ratio, and it could be difficult to match the low loop eccentricity.

Possibly deeper exposures on larger scales of the NGC5907 halo could 
reveal many very faint new structures that may help disentangle
 between these two alternatives.  Using state-of-the-art AREPO 
code to reproduce the NGC5907 loops would also be useful for providing
 a more realistic the disc and the distribution of matter in its surrounding.


\subsection{Concluding remarks}
 
This study shows that a major merger scenario for the formation of giant
loops in NGC 5907 may explain the observations, 
as well as the infall of a minor merger with a mass ratio lower than 12:1.
 Together with the rebuilding of a thin disc in an almost bulge-less galaxy, a gas-rich
major merger forms gigantic loops, within which stars may orbit
in loops for several billion years. Low-mass ratios down to three are
particularly efficient in removing the residuals of the secondary
interloper, and are still consistent with the observational
uncertainties on the bulge-to-total mass ratio. This is especially
true when high order loops are considered, several billion years
after the merger. This contrasts with minor mergers
that are assumed to orbit at large distance from the galaxy centre,
and should have left imprints of a residual core. This proposition of a
major merger origin could apply as well to the faint tidal features
found in many other spiral galaxies. Interestingly, our model of NGC 5907 can be falsified by new and
extremely deep observations of a wider field surrounding this galaxy: if
NGC 5907 has been formed by a major merger, there should be faint
and extremely large structures at hundred kilo-parsecs or more,
coinciding with lower order loops.

Future work will include modelling other nearby spiral galaxies with large and faint, extended 
features in their haloes. Most of them show no trace of a residual core of a minor merger 
residual. While it is accepted more and more that bulge-less galaxies could result from gas-rich 
major mergers \citep{guedes2011,brook2011,font2011}, the consecutive modelling of faint features 
in their halo should put considerable constraints on the orbital parameters. A wide diversity 
of orbital parameters has been already found by \citet{hammer2009} in modelling distant, unrelaxed 
galaxies by major mergers. These distant galaxies are likely similar to the progenitors, six billion 
years ago, of present-day spirals, and linking them together could provide another crucial test 
for the spiral rebuilding disc scenario.

\begin{acknowledgements}
This work has been supported by the ''Laboratoire International Associ\'e'' Origins, and computations were done using the special supercomputer at the Center of Information and Computing at National Astronomical Observatories, Chinese Academy of Sciences, funded by Ministry of Finance under the grant ZDYZ2008-2, as well as at the Computing Center at the Paris Observatory. We are very grateful to Sylvestre Taburet and the GEPI software team for their precious help.  The images in gas surface density of Figs 1, 3, 4, and the bottom
row of Fig. 5 were produced using SPLASH \citep{price2007}. 

\end{acknowledgements}

\appendix
\section{Implementation of cooling, feedback and star formation in GADGET-2}

To account for cooling, star formation, and feedback processes of the
ISM, we modified the publicly available version of the GADGET-2 code by adding these additional processes. 
We followed the method of \citet{cox2006} and \citet{springel2000} to 
implement these physical processes. We give a brief description 
below and details are described in \citet{cox2006}. 

Radiative cooling is important for gas to cool down to the central
region of the dark halo and then form stars. The cooling 
rate is calculated following \citet{katz1996}, in
which the primordial plasma of H and He and collisional 
ionisation equilibrium are assumed. Gas particles in these 
dense regions can form stars. The star formation rate is 
correlated to local gas density and anti-correlated to the local dynamic time: 

\begin{equation}
\frac{{\rm d} \rho_\star} {{\rm d}t} = C_{\star}\frac{\rho_{gas}}{t_{dynamic}}
\end{equation}

where 

\begin{equation}
t_{dynamic}=(4{\pi}G\rho_{gas})^{-\frac{1}{2}} .
\end{equation}

The method of determining star formation rate relies on the
Kennicutt-Schmidt law \citep{kinnicutt1998} according to which
$\frac{{\rm d}\rho_{\star}}{{\rm d}t} \propto \rho_{gas}^{1.5}$.
Gas particles are converted into collisionless particles according to the
above equation, using a stochastic technique.

To regulate star formation, feedback is needed to prevent gas from
being consumed by star formation. It assumes that the energy released
by a supernova is first stored in a new reservoir of internal energy. 
The energy from this new reservoir will provide additional pressure to
support the gas and prevent it from further collapsing to form stars. 
The feedback energy $q$ can also be thermalised. \citet{cox2006}  
introduced two free parameters to control this thermalisation process. 
One gives the time scale $\tau_{feedback}$ of thermalisation process, 
the other controls the equation of state.

We verified that our implementation of additional routines
that track the radiative cooling of gas and star formation.in the GADGET-2 code 
precisely follows that of \citet{cox2006}. We have indeed reproduced all the 
Cox et al figures for all their adopted values of feedback (low, medium, and high) 
and for the density dependence of the feedback energy thermalisation timescale (n= 0, 1 and 2)..

\begin{figure}
\centering
\includegraphics[width=7cm]{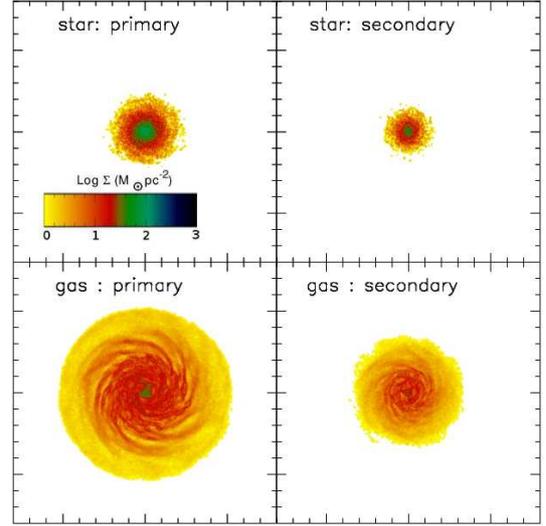}
\caption{Projected mass density of the primary (left) and secondary (right) galaxies 
for model with mass ratio 3 (model M3L34) for which the galaxies are simulated in 
isolation for $\sim$ 3 Gyr. The top row shows the stars and the bottom row show 
the gas distribution. A colour bar indicates the surface mass density scale shown in 
the top-left panel. The size of each panel is 160 by 160 kpc. }
\label{fig:relaxyimg}
\end{figure}

\begin{figure}
\centering
\includegraphics[width=7cm]{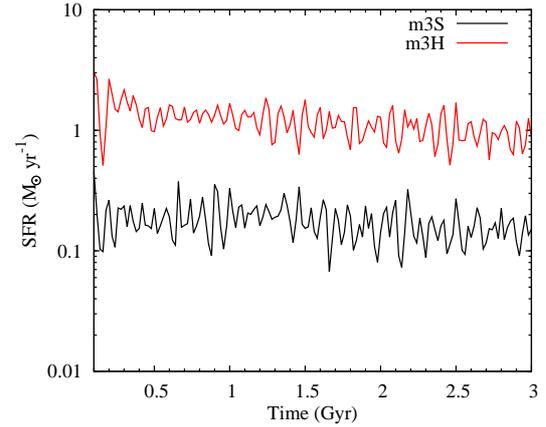}
\caption{Star formation rates for isolated galaxy models of the
  primary (top curve) and the 
secondary (bottom curve) for a mass ratio 3 (model M3L34). Five times the median 
feedback of \citet{cox2006} is used in these models.}
\label{fig:relaxysfr}
\end{figure}

We also tested the stability of our model galaxies by running them
in isolation as previously done by \citet{cox2006}. 
This is shown in Fig. \ref{fig:relaxyimg}, which shows 
how the star and gas surface mass density have evolved after 3 Gyr. The gas 
and star components for both primary and secondary
galaxies are stable during isolated evolution.  
This stability is also reflected by the evolution of the star 
formation rate as shown in Fig. \ref{fig:relaxysfr}.  During 
the 3 Gyr evolution, the star formation keeps a roughly constant value. 
The star formation rate of the secondary is about 0.18 M$_{\odot}$ yr$^{-1}$ and it is $\sim$ 1.1 M$_{\odot}$ yr$^{-1}$ for the primary galaxy. 
It slowly decreases because the star formation affects the gas surface density.
The star formation rate curve is not smooth because of the high
feedback value 
\citep[five times the median feedback of][]{cox2006}
that is 
adopted to regulate the star formation. This test shows that the current feedback model can lead  
to a stable disc and preserve the gas before fusion, which is needed
to form a disc after a merger.

\section{Properties of loops formed by particles coming back from tidal tails}
\label{sec:tailloop}

\citet{hammer2010} modelled the M31 galaxy with a  3:1, major merger event, which also 
reproduced the Giant Stream and, for some models, also the northwest big loop discovered 
by the PANDA team \citep{richardson2011}. In these models, particles returning to the 
galaxy from the tidal tail can feed the loops for several Gyrs ( $>$ 5 Gyr, see their Figure 8). 
Here we explore the properties of such loops, and trace the motion of tidal tail 
particles. In this section, we take an overview of the loop properties as they result from our numerical computations. 

Particles stripped into a tidal tail are of two kinds: part of them have enough energy 
to escape the galaxy potential, while others reach a maximum distance and then fall back 
to the galaxy. We are mostly concerned by the orbital pattern shape of particles
coming back from the tidal tail, because they can generate many faint features after 
the fusion. These features are very likely observable, because the process can be maintained 
for several Gyrs \citep[see e.g.][]{hammer2010}. These time estimates are considerably 
larger than those obtained for encounters of two ellipticals and in much better agreement
 with encounters of an elliptical and a spiral \citep{feldmann2008}. This argues for the 
necessity of a cold component in order to produce strong and long-lived tidal features.


A tidal tail formed in a major merger is more complex than one formed in a
minor merger. As shown by \citet{choi2007}, a massive 
satellite can change the morphology and radial velocity of a tidal
tail by self-gravity. Meanwhile, the satellite will also feel the
gravitational attraction from the tidal tail, affecting its energy and
angular momentum. The resonance may affect the tail properties \citep{donghia2010}. 
All these effects make tidal tail formation in major
mergers more complex than in minor mergers and thus more difficult to
reproduce. Important properties of the orbital pattern of tidal tails  
include the eccentricity of loops, which are approximated by 
ellipses, the width of the loops, and the precession rate of 
the particle orbits. 

In the following we describe the interaction between the most massive
(the primary) and the less massive (the secondary) galaxy in a
major merger. To match the NGC 5907 loops better, we need to
form two tidal tails. In the classic case of an encounter between two
prograde discs, we 
witness the formation of two tidal tails, one from each galaxy.
On the other hand, in our model one galaxy rotates prograde and the
other retrograde so that both tails form from material stripped from the
prograde companion, which in our model is the less massive one. The
two tails form at different times, namely the times of the first and
the second  closest passages.    
The motion of all the particles that constitute the tidal tail
is set before the interaction by the orbit of the secondary and its disc
spin. After being stripped from the secondary by the gravitational
interaction, these particles keep some orbital information from their
initial orbits, i.e. from their progenitors. When the secondary is
close to pericentre where  the tidal perturbation is maximum, the tidal 
force  will add energy to the particles  of the secondary \citep{sw1999}, 
which makes its particles unbound and ejected \citep{donghia2010}.  
Along the tidal tail, particle binding energy increases with radius \citep[see][]{hm1995,dubinski1996}, 
i.e., particles ejected further from the centre have higher energy, leading to an increasing fallback time.

\begin{figure*}
\centering
\includegraphics[width=18cm]{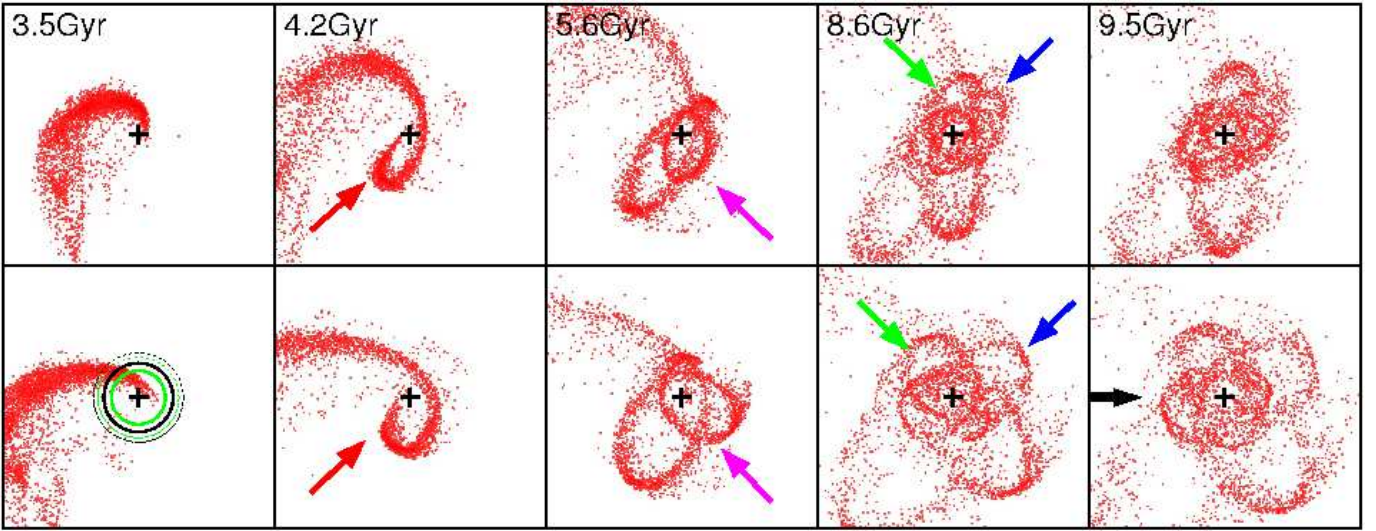}
\caption{Loops formed by the particles falling back from second tidal
  tail. Each panel is centred on the centre of mass of the secondary
  galaxy, indicated by a cross.  
In this figure, the tidal tail is formed from a 3:1 merger (model
M3L34, see section \ref{sec:result}).  The size of
each panel is 300 by 300 kpc. The stellar mass surface density of the
loops at 8.6 Gyr is consistent with the estimation of 
\citet{m2008} for NGC 5907. The top row
shows the loop formation process as seen from the observed viewing 
angle, while the bottom row gives a face-on view of the loops at the
same times. The red
arrows in the second column indicate the first loop, the 
pink arrows in the third column show the second loop, the blue and
green arrows in the fourth column show the third and fourth loops, and
the black arrow in the last column shows the fifth loop. In the 
bottom-left panel, solid and dotted circles indicate the radii of the
original location of the particles which constitute 
the tidal tails at 5.6 Gyr and 8.6 Gyr, respectively. 
Particles with small elevations are selected from the region between
the green and  
the black circles, and those with high elevation, from beyond the black circle.}
\label{fig:loops}
\end{figure*}

The motion of particles falling back from the tidal tail can be easily
understood since the dynamical friction force can be neglected due to
their low mass. If we also neglect the time evolution of the central
potential, the motion is simply that of test particles in a static central
potential, which is a rosette (Binney \& Tremaine 1987, hereafter
BT87). The angle between two consecutive apocentra depends on the
potential, being 0 degrees in a Kepler potential and 180 degrees in a
harmonic potential. Thus the orbit can be approximated by a precessing
ellipse, while the precession angle can give information on the
potential of the merger remnant.

\begin{figure*}
\centering
\includegraphics[width=7cm]{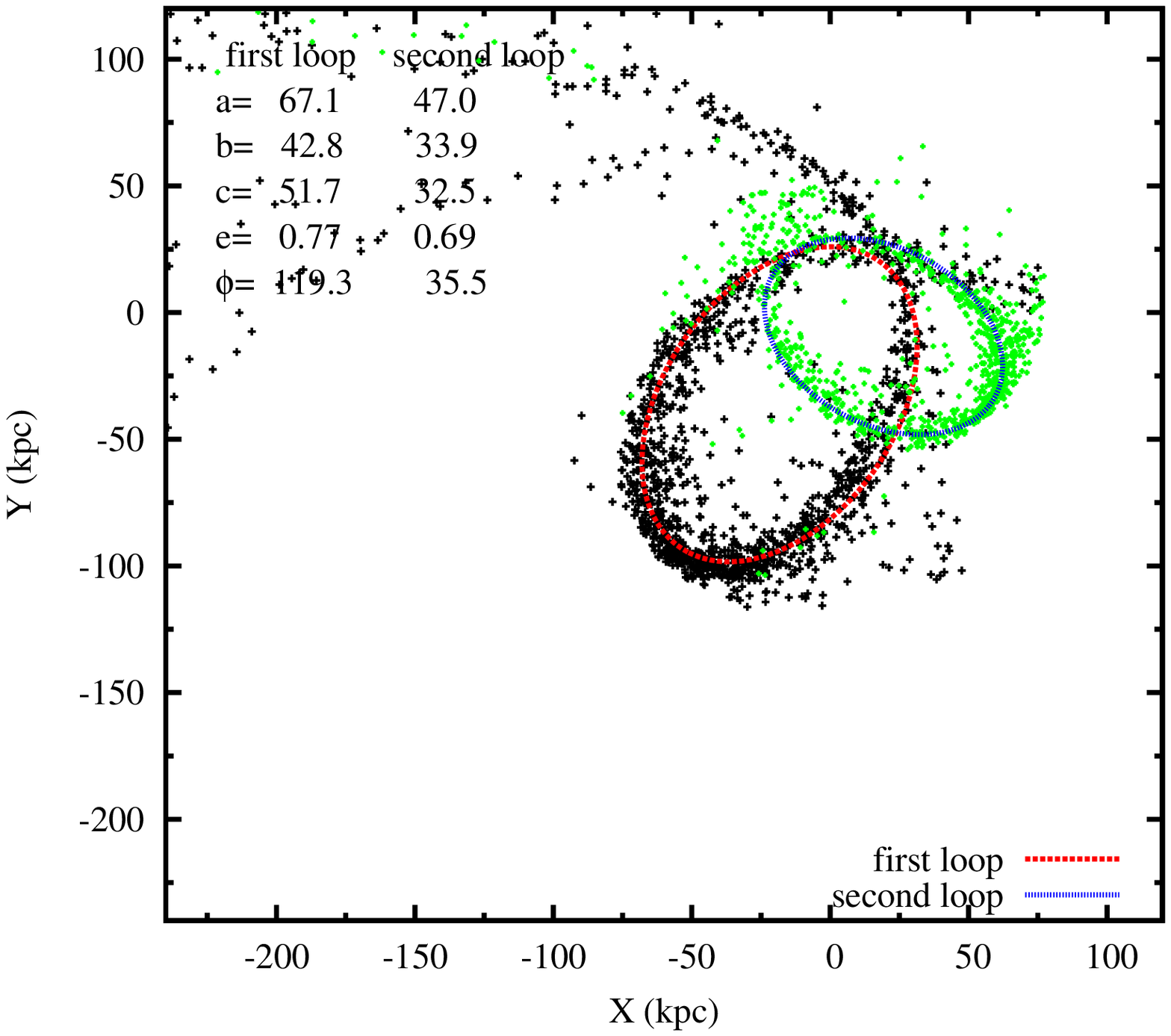}
\includegraphics[width=7cm]{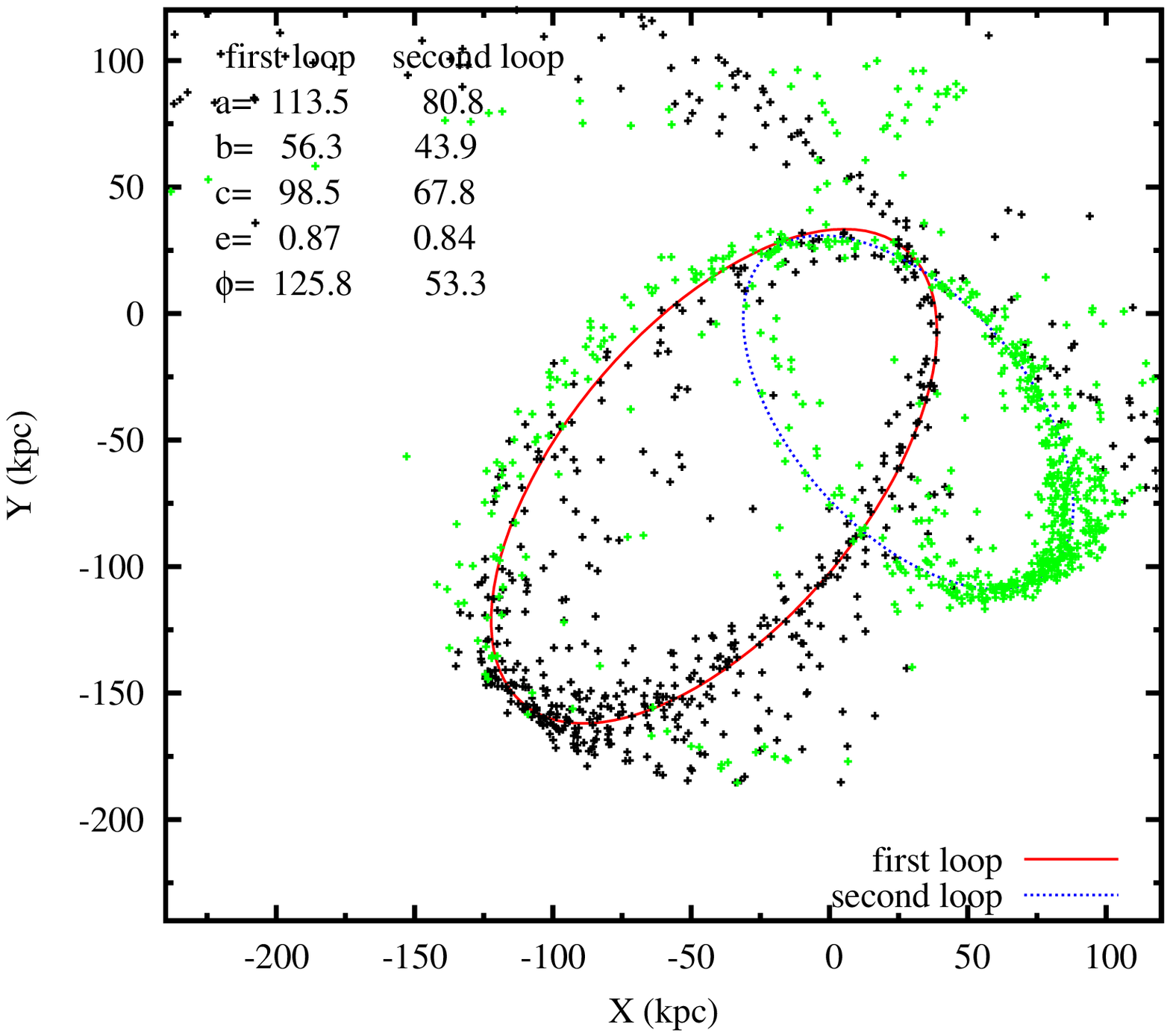}
\caption{First and second loops for the same model as in Fig. \ref{fig:loops} are fitted by ellipses for simulation with 
mass ratio 3. The left panel is at 5.6 Gyr and the right panel shows loops at 8.6 Gyr. Particles belonging to different 
loops are naturally selected by their distance to the centre in the tail-forming epoch (3.5 Gyr). Green indicates particles 
with small elevations (see Fig. \ref{fig:loops}, bottom-left panel), and black points indicate large elevation particles. 
Green points populate the second loop, while black points show particles of the first loop. Particles coming back from 
the tidal tail begin to outline the first loop and then enter in the second loop. The parameters of the fitted ellipses are 
given in the top left corner of each panel. They are the semi-major axis a, semi-minor axis b, focus distance, 
eccentricity, and position angle.}
\label{fig:fitloop}
\end{figure*}

Figure \ref{fig:loops} shows an example of the particle locations at
different epochs. The tidal tail particles have been tagged at   
3.5 Gyr after the beginning of the simulation, i.e. after the fusion
of the two galaxies  (see Sec. \ref{sec:result}), so that we can
follow their individual motions. 
These particles continue to feed the loops as they return from the tidal
tail. At 4.2 Gyr the first loop is already formed, and then at 5.6 Gyr a second 
loop is added. As time increases further, the third and fourth loops are formed,
as shown by the 8.6 Gyr panel of Fig. \ref{fig:loops}. 

\begin{figure*}
\centering
\includegraphics[width=7cm]{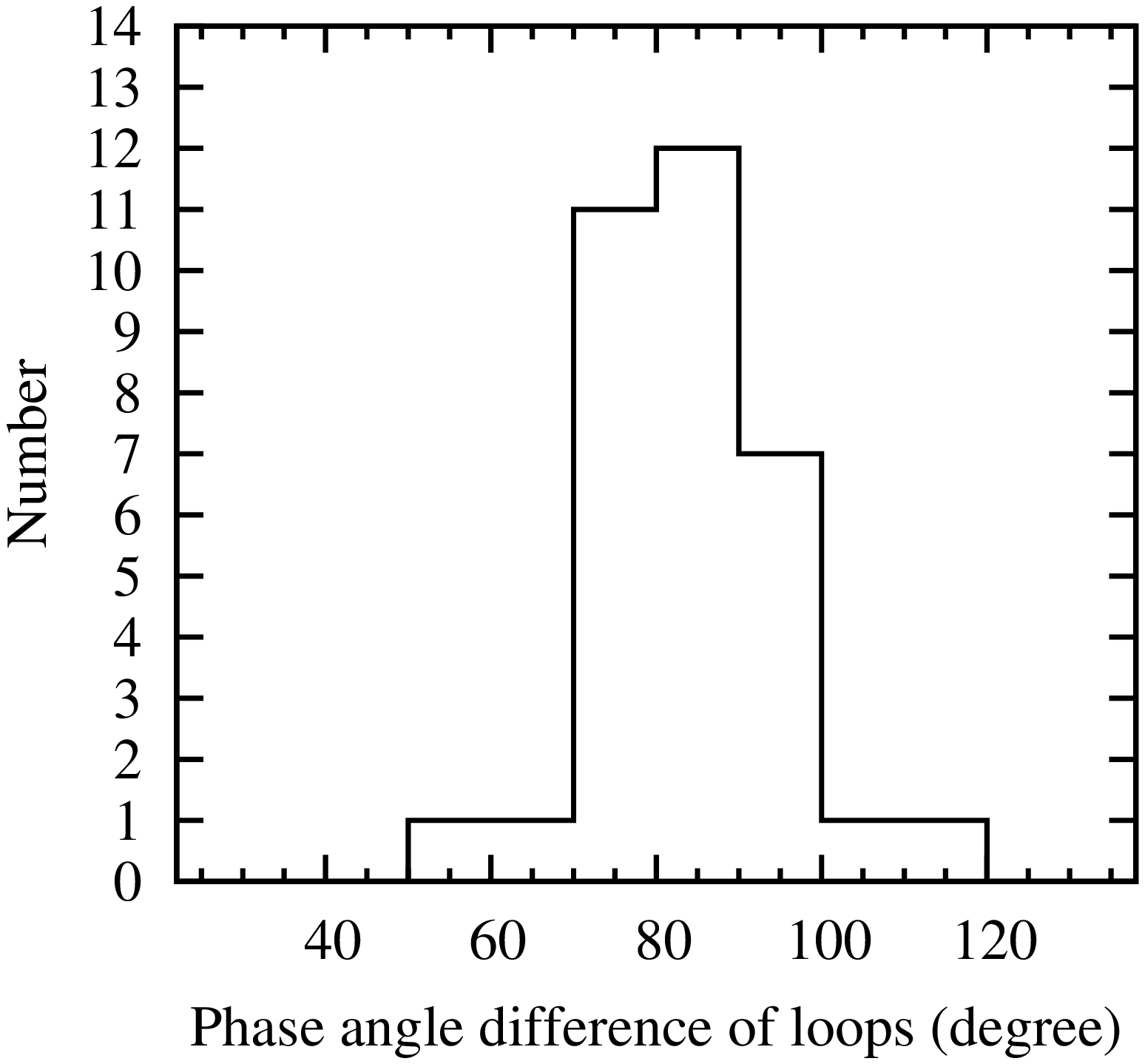}
\includegraphics[width=7cm]{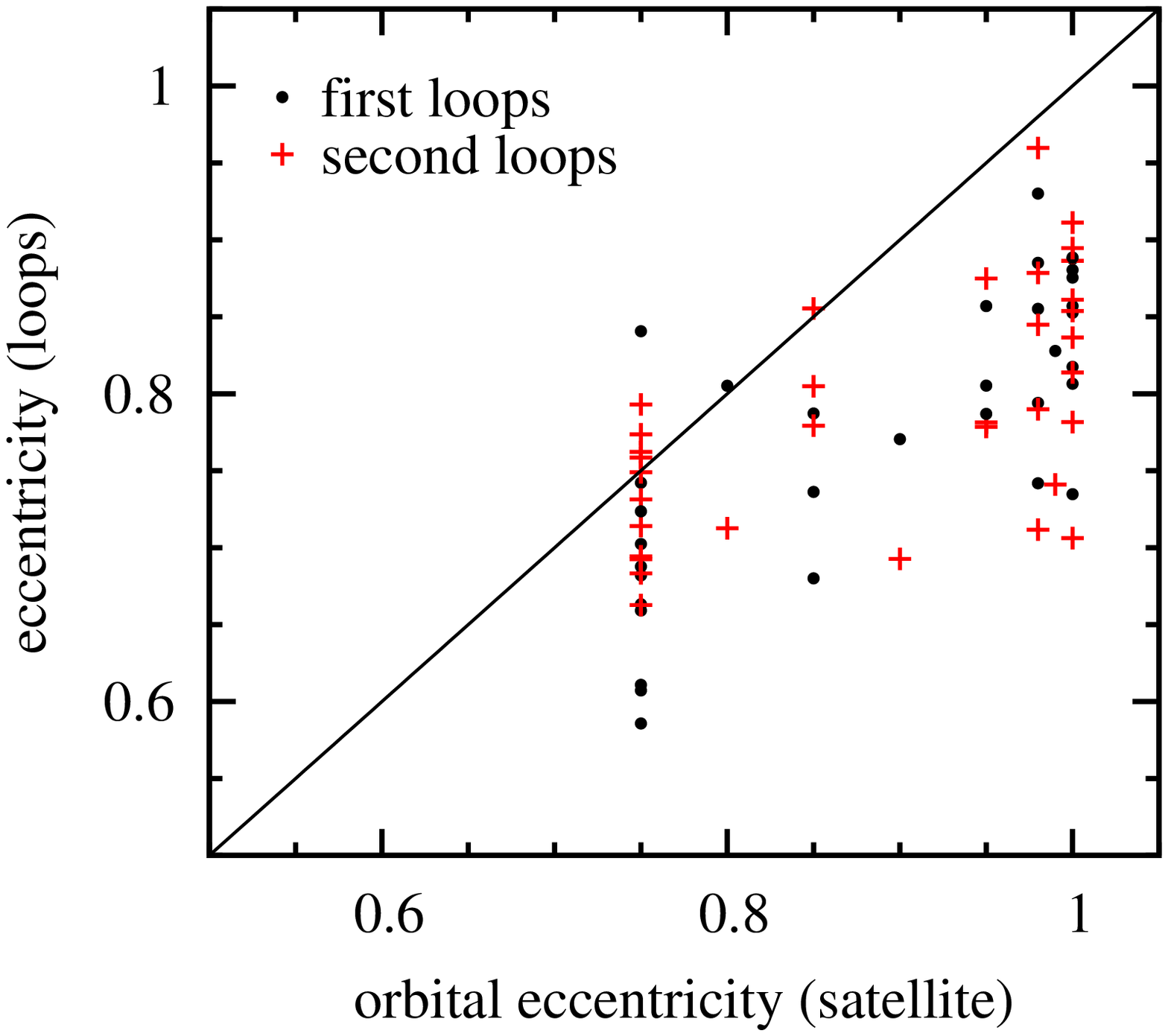}
\caption{The left panel gives the distribution of loop phase angle differences between first and second loops.
The right panel compares the loop and orbit eccentricities. Solid black line is the one-to-one relation. 
This statistics is coming from various models with different mass ratio (3, 4, and 5), orbital geometry and dark matter fraction.}
\label{fig:statloop}
\end{figure*}

Particles are continuously feeding the loops following a similar track from the 
tidal tail. They first enter into the first loop and then continue by orbiting in 
higher order loops. Because particles falling back later have higher energies 
(or elevations relatively to the mass centre), the size of the first loop increases with 
time, as well as that of other loops after their formation (see Fig. \ref{fig:loops}).
We also fitted the first and second loops with ellipses at 5.6 Gyr and 8.6 Gyr 
as shown in Fig. \ref{fig:fitloop}. Owing to the permanent motion of particles, 
it is not always simple to separate particles from different loops. We used 
the distance of particles from the centre at the tail-forming time (3.5 Gyr panel of
Fig. \ref{fig:loops}) to isolate particles for each loop.
In Fig. \ref{fig:fitloop} different colours indicate different
distances to the centre. The green particles are initially,
at the time of forming the tail, closer to the centre and now form the second loop, 
while the black particles that are farther from the centre form the
first loop. This shows that the particles forming the different loops come
from different distances to the centre 
at the time the tidal tail is formed. In this simulation we notice
that the first two loops formed by particles returning after reaching
their apocentra are 
approximated well by two ellipses with a phase angle $\sim$ 80 degree
difference due to precession, as shown in the left hand panel of
Fig. \ref{fig:statloop}.

The loop width depends on the orbital properties of the tidal tail. It
has been shown that resonance is important for forming a long tidal tail 
for a spin system. Strong resonance occurs when rotation and orbital
frequencies  obey  $\Omega_{disc} = \Omega_{orbit}$ \citep{tt1972,bh1992}. 
As long as we can 
control the tidal tail length and width, we may control the shapes of
the loops at some level. For various merger geometries, we can 
decrease the loop eccentricities by decreasing the orbital
eccentricity of the secondary, because they are related. This is
illustrated in the right hand 
panel of Fig. \ref{fig:statloop}, where we compare the loop eccentricity
with the orbital eccentricity of the second-order loop. Loops used in 
this figure are coming from models with different mass ratio, pericentre, eccentricity, baryonic fraction (from 6\% to 9\%), and 
gas fraction. The arrow indicates the region for which orbital eccentricities provide loop eccentricities consistent with the observations.


\begin{thebibliography}{}
\bibitem[Ashby et al.(2004)]{ashby04} Ashby, M.~L.~N., Pipher, J.~L., Forrest, W.~J., et al.\ 2004, Bulletin of the American Astronomical Society, 36, 1444
\bibitem[Barnaby \& Thronson (1992)]{barnaby1992} Barnaby, D., \& Thronson, H.~A., Jr. 1992, \aj, 103, 41 
\bibitem[Barnes(1992)]{barnes1992} Barnes, J.~E. 1992, \apj, 393, 484 
\bibitem[Barnes(2002)]{barnes2002} Barnes, J.~E. 2002, \mnras, 333, 481 
\bibitem[Barnes \& Hernquist(1992)]{bh1992} Barnes, J.~E., \& Hernquist, L.\ 1992, \araa, 30, 705 
\bibitem[Barnes \& Hernquist(1996)]{bh1996} Barnes, J.~E., \& Hernquist, L. 1996, \apj, 471, 115 
\bibitem[Bell et al.(2003)]{bell2003} Bell, E.~F., McIntosh, D.~H., Katz, N., \& Weinberg, M.~D. 2003, \apjs, 149, 289
\bibitem[Bell et al.(2006)]{bell2006} Bell, E.~F., Phleps, S., Somerville, R.~S., Wolf, C., Borch, A., \& Meisenheimer, K. 2006, \apj, 652, 270 
\bibitem[Benjamin et al.(2007)]{benjamin2007} Benjamin, R.~A., Draine, B.~T., Indebetouw, R., et al.\ 2007, The Science Opportunities of the Warm Spitzer Mission Workshop, 943, 101 
\bibitem[Bessell(2005)]{bessell2005} Bessell, M.~S. 2005, \araa, 43, 293
\bibitem[Binney \& Tremaine(1987)]{bt1987} Binney, J., \& Tremaine, S. 1987, Princeton, NJ, Princeton University Press, 1987 (BT87) 
\bibitem[Bizyaev \& Mitronova(2002)]{bizyaev2002} Bizyaev, D., \& Mitronova, S. 2002, \aap, 389, 795 
\bibitem[Bournaud et al.(2011)]{bournaud11} Bournaud, F., Chapon, D., Teyssier, R. et al., \apj, 730, 4 
\bibitem[Bridge et al.(2007)]{bridge2007} Bridge, C.~R., et al. 2007, \apj, 659, 931 
\bibitem[Brook et al.(2011)]{brook2011} Brook, C.~B., Stinson, G., Gibson, B.~K., et al.\ 2011, arXiv:1105.2562 
\bibitem[Burgdorf et al.(2009)]{burgdorf2009} Burgdorf, M., Ashby, M.~L.~N., Pang, S., \& Gilmore, G.~F.\ 2009, The Evolving ISM in the Milky Way and Nearby Galaxies,  
\bibitem[Choi et al.(2007)]{choi2007} Choi, J.-H., Weinberg, M.~D., \& Katz, N. 2007, \mnras, 381, 987 
\bibitem[Comeron et al.(2011)]{comeron2011} Comeron, S. et al., N. 2011, \apj, 729, 18 
\bibitem[Cox et al.(2006)]{cox2006} Cox, T.~J., Jonsson, P., Primack, J.~R., \& Somerville, R.~S. 2006, \mnras, 373, 1013 
\bibitem[Daddi et al.(2010)]{daddi2010} Daddi, E., Bournaud, F., Walter, F., et al. 2010, \apj, 713, 686 
\bibitem[D'Onghia et al.(2009)]{donghia2009} D'Onghia, E., Besla, G., Cox, T.~J., \& Hernquist, L. 2009, \nat, 460, 605 
\bibitem[D'Onghia et al.(2010)]{donghia2010} D'Onghia, E., Vogelsberger, M., Faucher-Giguere, C.-A., \& Hernquist, L. 2010, \apj, 725, 353 
\bibitem[Dubinski et al.(1996)]{dubinski1996} Dubinski, J., Mihos, J.~C., \& Hernquist, L. 1996, \apj, 462, 576 
\bibitem[Dubinski et al.(1999)]{dubinski1999} Dubinski, J., Mihos, J.~C., \& Hernquist, L. 1999, \apj, 526, 607 
\bibitem[Dutton et al.(2010)]{dutton2010} Dutton, A.~A., Conroy, C., van den Bosch, F.~C., Prada, F., \& More, S. 2010, \mnras, 407, 2
\bibitem[Elmegreen et al.(2009)]{elmegreen2009} Elmegreen, B. G., Elmegreen, D. M., Fernandez, M. X. \& Lemonias, J. J. 2009, \apj, 692, 12
\bibitem[Erb et al.(2006)]{erb2006} Erb, D.~K., Steidel, C.~C., Shapley, A.~E., et al. 2006, \apj, 646, 107 
\bibitem[Fakhouri \& Ma(2008)]{fm2008} Fakhouri, O., \& Ma, C.-P. 2008, \mnras, 386, 577 
\bibitem[Feldmann et al. (2008)]{feldmann2008} Feldmann, R., Mayer, L. \& Carollo, C. M.,  2008, \apj, 684, 1062
\bibitem[Font et al.(2011)]{font2011} Font, A. S. et al., 2011, \mnras submitted (arXiv:1103.0024) 
\bibitem[Garcia-Burillo et al.(1997)]{garcia1997} Garcia-Burillo, S., Guelin, M., \& Neininger, N. 1997, \aap, 319, 450 
\bibitem[Guedes et al.(2011)]{guedes2011} Guedes, J., Callegari, S., Madau, P., \& Mayer, L.\ 2011, arXiv:1103.6030 
\bibitem[Haan et al.(2011)]{haan2011} Haan, S., Surace, J.~A., Armus, L., et al.\ 2011, \aj, 141, 100           
\bibitem[Hammer et al.(2005)]{hammer2005} Hammer, F., Flores, H., Elbaz, D., Zheng, X.~Z., Liang, Y.~C., \& Cesarsky, C. 2005, \aap, 430, 115 
\bibitem[Hammer et al.(2007)]{hammer2007} Hammer, F., Puech, M., Chemin, L., Flores, H., \& Lehnert, M.~D. 2007, \apj, 662, 322 
\bibitem[Hammer et al.(2009)]{hammer2009} Hammer, F., Flores, H., Puech, M., Yang, Y.~B., Athanassoula, E., Rodrigues, M., \& Delgado, R.\ 2009, \aap, 507, 1313 
\bibitem[Hammer et al.(2010)]{hammer2010} Hammer, F., Yang, Y.~B., Wang, J.~L., Puech, M., Flores, H., \& Fouquet, S. 2010, \apj, 725, 542 
\bibitem[Hernquist(1990)]{hernquist1990} Hernquist, L.\ 1990, \apj, 356, 359 
\bibitem[Hibbard \& Mihos(1995)]{hm1995} Hibbard, J.~E., \& Mihos, J.~C. 1995, \aj, 110, 140 
\bibitem[Hoekstra et al.(2005)]{hoekstra2005} Hoekstra, H., Hsieh, B.~C., Yee, H.~K.~C., Lin, H., \& Gladders, M.~D. 2005, \apj, 635, 73 
\bibitem[Hopkins et al.(2008)]{hopkins2008} Hopkins, P.~F., Hernquist, L., Cox, T.~J., Younger, J.~D., \& Besla, G. 2008, \apj, 688, 757 
\bibitem[Hopkins et al.(2009)]{hopkins2009} Hopkins, P.~F., Cox, T.~J., Younger, J.~D., \& Hernquist, L. 2009, \apj, 691, 1168 
\bibitem[Hopkins et al.(2010)]{hopkins2010} Hopkins, P.~F., et al. 2010, \apj, 715, 202
\bibitem[House et al.(2011)]{house2011} House, E., et al. 2011, arXiv:1104.2037 
\bibitem[Ibata et al.(2001)]{ibata2001} Ibata, R., Irwin, M., Lewis, G., Ferguson, A.~M.~N., \& Tanvir, N. 2001, \nat, 412, 49 
\bibitem[Irwin \& Madden(2006)]{im2006} Irwin, J.~A., \& Madden, S.~C. 2006, \aap, 445, 123
\bibitem[Sesar(2011)]{sesar2011} Sesar, B.\ 2011, RR Lyrae Stars, Metal-Poor Stars, and the Galaxy, 135 (arXiv:1105.4146)
\bibitem[Just et al.(2006)]{just2006} Just, A., M{\"o}llenhoff, C., \& Borch, A. 2006, \aap, 459, 703 
\bibitem[Katz et al.(1996)]{katz1996} Katz, N., Weinberg, D.~H., \& Hernquist, L. 1996, \apjs, 105, 19
\bibitem[Kennicutt(1998)]{kinnicutt1998} Kennicutt, R.~C., Jr. 1998, \apj, 498, 541
\bibitem[Keres et al.(2011)]{keres2011} Keres, D., Vogelsberger, M., Sijacki, D., Springel, V., \& Hernquist, L. 2011, arXiv:1109.4638 
\bibitem[Khochfar \& Burkert(2006)]{kb2006} Khochfar, S., \& Burkert, A. 2006, \aap, 445, 403 
\bibitem[Kormendy \& Kennicutt(2004)]{kk2004} Kormendy, J., \& Kennicutt, R.~C., Jr.\ 2004, \araa, 42, 603 
\bibitem[Laine et al.(2010)]{laine2010} Laine, S., Appleton, P.~N., Gottesman, S.~T., Ashby, M.~L.~N., \& Garland, C.~A.\ 2010, \aj, 140, 753 
\bibitem[Lequeux et al.(1998)]{lequeux1998} Lequeux, J., Combes, F., Dantel-Fort, M., Cuillandre, J.-C., Fort, B., \& Mellier, Y. 1998, \aap, 334, L9
\bibitem[Lin et al.(2004)]{lin2004} Lin, L., et al. 2004, \apjl, 617, L9 
\bibitem[Lin et al.(2008)]{lin2008} Lin, L., et al. 2008, \apj, 681, 232 
\bibitem[Lotz et al.(2008)]{lotz2008} Lotz, J.~M., et al. 2008, \apj, 672, 177
\bibitem[Mart{\'{\i}}nez-Delgado et al.(2008)]{m2008} Mart{\'{\i}}nez-Delgado, D., Pe{\~n}arrubia, J., Gabany, R.~J., Trujillo, I., Majewski, S.~R., \& Pohlen, M. 2008, \apj, 689, 184 (M08)
\bibitem[Mart{\'{\i}}nez-Delgado et al.(2010)]{m2010} Mart{\'{\i}}nez-Delgado, D., et al. 2010, \aj, 140, 962 
\bibitem[Martig \& Bournaud(2010)]{mb2010} Martig, M., \& Bournaud, F. 2010, \apjl, 714, L275 
\bibitem[McGaugh (2005)]{mcgaugh2005} McGaugh, S.~S. 2005, \apj, 632, 859 
\bibitem[Miller \& Rubin(1995)]{mr1995} Miller, B.~W., \& Rubin, V.~C. 1995, \aj, 110, 2692 
\bibitem[Mouhcine et al.(2005)]{mouhcine2005} Mouhcine, M., Ferguson, H.~C., Rich, R.~M., Brown, T.~M., \& Smith, T.~E.\ 2005, \apj, 633, 821 
\bibitem[Peng et al.(2010)]{peng2010} Peng, C.~Y., Ho, L.~C., Impey, C.~D., \& Rix, H.-W. 2010, \aj, 139, 2097 
\bibitem[Price(2007)]{price2007} Price, D.~J. 2007, \pasa, 24, 159
\bibitem[Puech et al.(2008)]{puech2008} Puech, M., et al. 2008, \aap, 484, 173 
\bibitem[Puech et al.(2010)]{puech2010} Puech, M., Hammer, F., Flores, H., Delgado-Serrano, R., Rodrigues, M., \& Yang, Y. 2010, \aap, 510, A68 
\bibitem[Puech et al.(2011)]{puech2011} Puech M. et al. 2011, \apj, submitted
\bibitem[Reshetnikov \& Sotnikova(2000)]{rs2000} Reshetnikov, V.~P., \& Sotnikova, N.~Y. 2000, Astronomy Letters, 26, 277 
\bibitem[Richardson et al.(2011)]{richardson2011} Richardson, J.~C., Irwin, M.~J., McConnachie, A.~W., et al.\ 2011, \apj, 732, 76 
\bibitem[Robertson et al.(2006)]{robertson2006} Robertson, B., Bullock, J.~S., Cox, T.~J., Di Matteo, T., Hernquist, L., Springel, V., \& Yoshida, N. 2006, \apj, 645, 986 
\bibitem[Sancisi \& van Albada(1987)]{sv1987} Sancisi, R., \& van Albada, T.~S. 1987, Dark matter in the universe, 117, 67
\bibitem[Saha et al.(2009)]{saha2009} Saha, K., de Jong, R., \& Holwerda, B.\ 2009, \mnras, 396, 409 
\bibitem[Sanders \& Mirabel(1996)]{sm1996} Sanders, D.~B., \& Mirabel, I.~F. 1996, \araa, 34, 749 
\bibitem[Shang et al.(1998)]{shang1998} Shang, Z., et al. 1998, \apjl, 504, L23 
\bibitem[Springel(2000)]{springel2000} Springel, V. 2000, \mnras, 312, 859 
\bibitem[Springel(2005)]{springel2005} Springel, V. 2005, \mnras, 364, 1105 
\bibitem[Springel et al.(2005)]{sdh2005} Springel, V., Di Matteo, T., \& Hernquist, L.\ 2005, \mnras, 361, 776 
\bibitem[Springel \& Hernquist(2002)]{sh2002} Springel, V., \& Hernquist, L. 2002, \mnras, 333, 649
\bibitem[Springel \& Hernquist(2003)]{sh2003} Springel, V., \& Hernquist, L. 2003, \mnras, 339, 289 
\bibitem[Springel \& Hernquist(2005)]{sh2005} Springel, V., \& Hernquist, L. 2005, \apjl, 622, L9 
\bibitem[Springel et al.(2001)]{springel2001} Springel, V., Yoshida, N., \& White, S.~D.~M. 2001, \na, 6, 79 
\bibitem[Springel \& White(1999)]{sw1999} Springel, V., \& White, S.~D.~M. 1999, \mnras, 307, 162 
\bibitem[Stark et al.(2009)]{stark2009} Stark, D.~V., McGaugh, S.~S., \& Swaters, R.~A. 2009, \aj, 138, 392
\bibitem[Stewart et al.(2009)]{stewart2009} Stewart, K.~R., Bullock, J.~S., Wechsler, R.~H., \& Maller, A.~H. 2009, \apj, 702, 307
\bibitem[Toomre \& Toomre(1972)]{tt1972} Toomre, A., \& Toomre, J.\ 1972, \apj, 178, 623 
\bibitem[Toomre(1977)]{toomre1977} Toomre, A.\ 1977, Evolution of Galaxies and Stellar Populations, 401 
\bibitem[Toth \& Ostriker(1992)]{to1992} Toth, G., \& Ostriker, J.~P. 1992, \apj, 389, 5
\bibitem[Zheng et al.(1999)]{zheng1999} Zheng, Z., et al.\ 1999, \aj, 117, 2757 
\bibitem[Zibetti et al.(2004)]{zibetti2004} Zibetti, S., White, S.~D.~M., \& Brinkmann, J. 2004, \mnras, 347, 556 
\bibitem[van der Kruit \& Searle(1982)]{vs1982} van der Kruit, P.~C., \& Searle, L. 1982, \aap, 110, 61 
\bibitem[van der Kruit(2007)]{van2007} van der Kruit, P.~C. 2007, \aap, 466, 883 
\bibitem[van Dokkum \& Conroy(2010)]{vc2010} van Dokkum, P.~G., \& Conroy, C. 2010, \nat, 468, 940 
\end{thebibliography}
\end{document}